\documentclass[]{mn2e}

\newif\ifAMStwofonts

\usepackage{graphicx}

\def\bib{\parskip=0pt\par\noindent\hangindent\parindent \parskip =2ex plus .5ex
    minus .1ex}
\newcommand{\lsim}{\raisebox{-0.13cm}{~\shortstack{$<$ \\[-0.07cm] $\sim$}}~}
\newcommand{\gsim}{\raisebox{-0.13cm}{~\shortstack{$>$ \\[-0.07cm] $\sim$}}~}

\title[Further constraints on the evolution of $K_s$-selected galaxies]{Further constraints on the evolution of $K_s$-selected galaxies in the GOODS/CDFS field}

\author[K.I. Caputi, R.J. McLure, J.S. Dunlop, M. Cirasuolo  and  A.M. Schael] 
{K.I. Caputi$^{1}$\thanks{Email addresses: kcaputi@ias.u-psud.fr; \newline 
 rjm, jsd, ciras, ams@roe.ac.uk},
 R.J. McLure$^{2 \, \star}$,
 J.S. Dunlop$^{2 \, \star}$,
 M. Cirasuolo$^{2 \, \star}$
 and A.M. Schael$^{2 \, \star}$ 
\\
$^{1}$Institut d'Astrophysique Spatiale, b\^at. 121, 
Universit\'e Paris-Sud, F-91405 Orsay Cedex, FRANCE
\\
$^{2}$Institute for Astronomy, University of Edinburgh, Royal Observatory,
       Edinburgh EH9 3HJ, Scotland, U.K.
       }  

\date{ }

\pubyear{2005}

\begin{document}

\maketitle

\label{firstpage}

\begin{abstract}

  We have selected and analysed the properties of a sample of 2905 $K_s<21.5$ galaxies in $\rm \sim 131 \, arcmin^2$ of the Great Observatories Origins Deep Survey (GOODS) Chandra Deep Field South (CDFS), to obtain further constraints on the evolution of $K_s$-selected galaxies with respect to the results already obtained in previous studies. We made use of the public deep multiwavelength imaging from the optical $B$ through the infrared (IR) $\rm 4.5 \, \mu m$ bands, in conjunction with available spectroscopic and COMBO17 data in the CDFS, to construct an optimised redshift catalogue for our galaxy sample. We computed the  $K_s$-band LF and determined that its characteristic magnitude has a substantial brightening and a decreasing total density from $z=0$ to $\langle z \rangle =2.5$.  We also analysed the colours and number density evolution of galaxies with different stellar masses.  Within our sample, and in contrast to what is observed for less massive systems,  the vast  majority ($\sim 85-90$\%) of the most massive ($M>2.5 \times 10^{11} \, M_\odot$) local galaxies appear to be in place before redshift $z \sim1$. Around $65-70$\% of the total assemble  between redshifts  $z=1$ and $z=3$ and most of them display extremely red colours, suggesting that plausible star formation in these very massive systems should mainly proceed in obscured, short-timescale bursts.  The remaining fraction (up to $\sim 20$\%) could be in place at even higher redshifts  $z=3-4$, pushing the first epoch of formation of massive galaxies beyond the limits of current near-IR surveys.

\end{abstract}

\begin{keywords}
galaxies: evolution -- galaxies: formation -- galaxies: high-redshift -- galaxies: luminosity function, mass function
\end{keywords}

\setcounter{figure}{0}

\section{Introduction}
\label{sec-intro}

\parskip=0pt

Deep multiwavelength surveys are progressively sheding light on the history of galaxy evolution since very high redshifts. From new surveys in the ultraviolet regime (e.g. Martin et al. 2005) to the latest IR, sub-millimetre and radio campaigns (e.g. Werner et al. 2004; Dunlop 2005; Condon et al. 2003), the advent of numerous datasets is enhancing our understanding of when galaxies were formed and how they evolved through cosmic time. UV/optical observations can unveil the sources of stellar emission but are limited to detect unobscured systems, missing an important fraction of the galaxies with on-going star formation. IR and longer-wavelength surveys provide an unbiased way of studying star-forming systems,  but are insensitive to galaxies with little or no dust content.  Near-IR observations, on the contrary, appear as one of the most suitable methods for making a complete census of galaxy populations, as they are sensitive to the stellar emission from both young and old galaxies, and are also relatively unaffected  by the presence of dust. The rest-frame $K$ band is the most efficient tracer of  the stellar mass content of the Universe.

 Recent studies of near-IR-selected galaxies allowed for a rapid progress in setting different constraints on the history of galaxy formation (Cimatti et al. 2002; Daddi et al. 2003; Caputi et al. 2004; Caputi et al. 2005a; among others) and, in particular, on the evolution of the stellar mass content of the Universe with redshift (Dickinson et al. 2003; Fontana et al. 2004; Glazebrook et al. 2004;  Caputi et al. 2005a; Drory et al. 2005).  It is now recognised that, although more than a half  of the stellar mass has been formed at redshifts $z \lsim 1.5$, near-IR-selected galaxies are already  present before at least redshifts $z \sim 2-3$  and contain a significant fraction of the stellar mass observed at redshift $z=0$.   However, one of the main problems of existing near-IR surveys is that, when sufficiently deep, they are usually restricted to small areas of the sky, making the results prone to large error bars due to cosmic variance. Also, the lack of statistics prevents refined studies of galaxy evolution, where trends for different near-IR sub-populations can be individually investigated.

 The GOODS project (Giavalisco et al.~2004) is a public program of multiwavelength observations which is now close to completion.  The data sets are kindly released to the astronomical community in a fully reduced mode. As part of the GOODS project, $~160 \, {\rm arcmin}^2$ of the CDFS have been observed with  the Advanced Camera for Surveys (ACS) on board the Hubble Space Telescope (HST) in four passbands: $B$, $V$, $I_{775}$ and $z_{850}$. In addition, the GOODS project includes deep near-IR imaging  taken with the Infrared Spectrometer and Array Camera (ISAAC) on the `Antu' Very Large Telescope (Antu-VLT). A first release of the European Southern Observatory (ESO) Imaging Survey (GOODS/EIS v0.5 release) provided data in the $J$, $H$ and $ K_s$ bands for $\sim 50 \,{\rm arcmin}^2$ of the GOODS/CDFS field. In this sub-region, Roche, Dunlop \& Almaini~(2003) selected a deep sample of Extremely Red Galaxies (ERGs) with  $ K_s \le 22$ (Vega), whose redshift distribution and luminosity evolution were studied by Caputi et al.~(2004). A detailed comparison of the properties of ERGs with other $K_s$-selected galaxies was presented by Caputi et al.~(2005a). Over the last year, however, new datasets have been made public within the GOODS project, increasing the coverage in wavelength and area of the GOODS/CDFS. The second GOODS/EIS (v1.0) release extends the deep $J$ and $ K_s$ band imaging to  $\rm \sim 131 \, arcmin^2$ of the GOODS/CDFS. Also, the Infrared Array Camera (IRAC) on board the Spitzer Space Telescope imaged this field at $3.6 \, {\rm \mu m}$,  $4.5 \, {\rm \mu m}$, $5.8 \, {\rm \mu m}$ and $8.0 \, {\rm \mu m}$. The availability of these data allows one to map the rest-frame $K_s$-band light up to redshifts $z \sim 2-3$.

 In this work, we take advantage of the further wavelength and area coverage of the GOODS/CDFS, in conjunction with the information  provided by the latest released spectroscopic and COMBO17 data, to set tighter constraints on the properties and evolution of $K_s$-selected galaxies. The layout of this paper is as follows. First, in Section \ref{sec_sample}, we present the sample selection and give details on the multiwavelength photometry.  In Section \ref{sec_redsh}, we explain the technique we applied to construct an optimised redshift catalogue for our galaxy sample and show the resulting redshift distribution. In Section \ref{sec_lf}, we present the rest-frame $K_s$-band luminosity function (LF) for our sample of $K_s$-selected galaxies. In Section \ref{sec_mass}, we discuss several aspects of the evolution of different mass systems. In Section \ref{sec_nearir}, we explore additional hints from the study of the purely near-IR colours of our galaxies. Finally, in Section \ref{sec_conc} we present some concluding remarks. We adopt throughout a cosmology with $\rm H_o=70 \,{\rm km \, s^{-1} Mpc^{-1}}$, $\rm \Omega_M=0.3$ and $\rm \Omega_\Lambda=0.7$. A Salpeter (1955) IMF over stellar masses $M=(0.1-100) \, M_\odot$ is assumed, unless where otherwise stated.

\section{The sample - multiwavelength photometry}
\label{sec_sample}

\parskip=0pt

 We used the public deep ISAAC $K_s$-band images in the GOODS/EIS v1.0 release to select a catalogue of $K_s<21.5$ (Vega) sources in 131 arcmin$^2$ of the GOODS/CDFS. This new release is composed of 21 maps which  cover more than twice the area of the GOODS/EIS v0.5 release. In addition, further improvements have been applied to the v1.0 imaging,  such as a revised data reduction technique and a uniform photometric calibration of the maps across the entire field.

 We performed the source extraction using the public code SEXTRACTOR (Bertin \& Arnouts, 1996) on the $K_s$-band images. We cut the resulting catalogue at a magnitude $K_s=21.5$, a limit half a magnitude brighter  than the cut we used to select our $K_s$ sample in Caputi et al. (2005a). Several reasons justify our new magnitude limit. First, a few of the additional maps in the v1.0 release have lower exposure times than the average exposure times of the maps composing the v0.5 release.  Second, this new limit constitutes an adequate balance between having access to galaxies up to very high redshifts  and requiring for our  catalogue to have a high level of completeness. The completeness levels of our v1.0 sample are very similar to those estimated for the v0.5 sample by Caputi et al. (2005a): 100\%, 90\% and 86\% for $K_s \leq 20.0$, $20.0 < K_s \leq 21.0$ and $21.0 < K_s \leq 21.5$, respectively. Finally, our new sample has been selected having in mind the possibility of a complete quantitative study of galaxy  morphology, only possible for high signal-to-noise objects. The results of the latter will be presented in a companion paper (McLure et al.~2005, in preparation). Our total v1.0 $K_s<21.5$ sample consists of 3184 sources.
 
 We measured multiwavelength photometry for our $K_s$-selected galaxies in order to model their spectral energy distribution (SED), as we shall explain in  Section \ref{sec_redsh}.  We ran SEXTRACTOR in `double-image mode' to perform aperture photometry on the GOODS/EIS v1.0 $J$-band images, centred at the position of the $K_s$-band extracted sources. We measured  magnitudes in an aperture size of 2.83-arcsec diameter for all the objects, to coincide with one of the aperture sizes used to construct the public GOODS/CDFS ACS catalogues. The latter have  been constructed running SEXTRACTOR on the $z_{850}$-band images, and making `double-image mode' runs to perform photometry in the $B$, $V$ and $I_{775}$ bands. We looked for counterparts of our $K_s<21.5$ (Vega) sources in the ACS catalogues and adopted the corresponding 2.83-arcsec diameter aperture magnitudes in each of the four ACS passbands. A 2.83-arcsec diameter aperture is sufficient to collect all the flux of the vast majority ($>$90\%) of our galaxies in the different optical/near-IR passbands. Even in the $K_s$ band, virtually all of our galaxies have 2.83-arcsec-diameter aperture magnitudes in agreement with the corresponding total $K_s$ magnitudes within the error bars. 
 
\begin{figure}
\begin{center}
\includegraphics[width=1.0\hsize,angle=0] {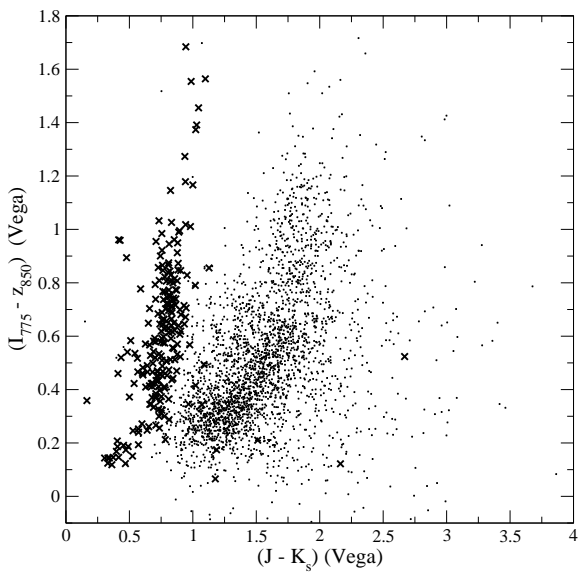}
\caption[]{\label{colstar} The ($I_{775}-z_{850}$) versus ($J-K_s$) colours for the sources in the $K_s<21.5$ sample. Cross-like symbols correspond to those objects with SEXTRACTOR stellarity parameter CLASS\_STAR$\geq$0.8.}
\end{center}  
\end{figure}

 Deep Spitzer/IRAC imaging for the GOODS/CDFS is now also publicly available, covering the wavelength range $\rm 3.6 \, \mu m$ to $\rm 8.0 \, \mu m$. At the time of writing, only the first epoch of IRAC GOODS/CDFS data had been released. This first epoch includes $\rm 3.6 \, \mu m$  and  $\rm 5.8 \, \mu m$ images for $\sim 2/3$ of the area of the GOODS/CDFS and $\rm 4.5 \, \mu m$  and  $\rm 8.0 \, \mu m$ images for another $\sim 2/3$ of this area, in such a way that only $\sim 1/3$ of the field have data in the four IRAC passbands. We ran SEXTRACTOR on the $\rm 3.6 \, \mu m$  and  $\rm 4.5 \, \mu m$  images and looked for counterparts of the $K_s$-band sources in a matching radius of 1.2 arcsec. We restricted the photometry to only the first two IRAC channels, as the templates we used for the galaxy SED fitting  lose accuracy or do not cover wavelengths beyond $\sim \rm 5 \, \mu m - 6 \, \mu m$.  We measured magnitudes in 2.83-arcsec diameter apertures. However, the point-spread function (PSF) of the IRAC images is much larger than the characteristic PSF of the ISAAC and ACS images: the mean full-width half maximum (FWHM) of the  IRAC  $\rm 3.6 \, \mu m$  PSF is 1.66 arcsec (Fazio et al. 2004), in comparison with 0.4 arcsec and 0.1 arcsec for the $K_s$ and $z_{850}$ bands in the ISAAC and ACS instruments, respectively. Thus, a 2.83-arcsec diameter aperture is insufficient to encompass the magnitudes measured in the IRAC bands with the magnitudes obtained in equivalent apertures on the ACS and ISAAC images. We computed aperture corrections for the magnitudes measured on the IRAC images studying the curve of growth of isolated stars in the field. We determined aperture corrections of 0.50 and 0.55 magnitudes for the 2.83-arcsec-diameter aperture magnitudes in the $\rm 3.6 \, \mu m$  and the $\rm 4.5 \, \mu m$  bands, respectively.

  To clean our catalogues for stars and active galaxies, we applied criteria somewhat more strict than those implemented by Caputi et al.~(2005a).  We cleaned our catalogues for stars using the SEXTRACTOR stellarity parameter `CLASS\_STAR', as measured on the  ACS $z_{850}$ images. To separate active galaxies, we looked for counterparts of our $K_s<21.5$ sources in the public X-ray catalogues available for the CDFS (Szokoly et al. 2004).
  
   222/3184 sources of our initial catalogue had CLASS\_STAR$\geq$0.8. Figure~\ref{colstar} shows the location of these 222 sources in a colour-color diagram, in comparison to the remaining sources in the catalogue. We see that the majority of the objects with CLASS\_STAR$\geq$0.8 occupy a clearly separated region in this plot, typically displaying colours $(J-K_s) \lsim 1.1$ (Vega). However, we find 6 outliers in the colour-colour diagram, which we studied on a case-by-case basis.  4 out of these 6 outliers are X-ray classified as active galactic nuclei (AGN) or quasi-stellar objects (QSO). The remaining 2 are not detected in the X-rays, but their SEDs cannot be satisfactorily fitted by HYPERZ (Bolzonella, Miralles \& Pell\'{o}, 2000) with any of the templates  in the GISSEL98 library (Bruzual \& Charlot 1993), so they are likely to be also stars or AGN/QSO. As a consequence, we excluded the 222 sources with CLASS\_STAR$\geq$0.8 from our sample.

    On the other hand, we found that 142/3184 of our $K_s$ sources were detected in the X-rays. However, in principle, the fact that a galaxy is classified as active in the X-rays does not necessarily imply that its optical-to-near-IR SED is dominated by an active nucleus. To determine which X-ray-detected $K_s$ sources had to be excluded from our catalogue, we modelled their SEDs using the templates in the GISSEL98 library of HYPERZ. The GISSEL98 library contains a wide range of synthetic SEDs based on  stellar spectra but no power-law SED template.  We determined that the SEDs  of 57/142 X-ray-detected $K_s$ sources could not be satisfactorily fitted by any template in the GISSEL98 library, so we excluded these 57 objects from our catalogue, as it is likely that their  optical/near-IR light is at least partially contaminated by an active nucleus component. 5 additional  X-ray-detected $K_s<21.5$ sources did not have either a satisfactory fit of their SEDs, but they already had been excluded on the basis of the stellarity parameter criterion.  In summary, our final $K_s<21.5$ source catalogue consisted of 2905 galaxies.

  Other no-X-ray-detected active galaxies could still be present in our final galaxy catalog. Another  characteristic allowing to identify active galaxies is the power-law shape of their SEDs from the UV/optical through the IR regimes. Thus, to investigate the presence of additional contaminating sources within our sample, we cross-correlated our galaxies with the catalogue of $\rm 24 \, \mu m$-selected IR power-law sources  in the CDFS constructed by Alonso-Herrero et al.~(2005). Within our sample, we found only 8 no-X-ray-detected  $K_s<21.5$ sources  included in the Alonso-Herrero et al. catalogue. We conclude, then, that the fraction of remaining plausibly-contaminating sources within our sample is basically negligible.

\section{An optimised redshift catalogue for $K_s$-selected galaxies in the GOODS/CDFS}
\label{sec_redsh}

\begin{figure}
\begin{center}
\includegraphics[width=1.0\hsize,angle=0] {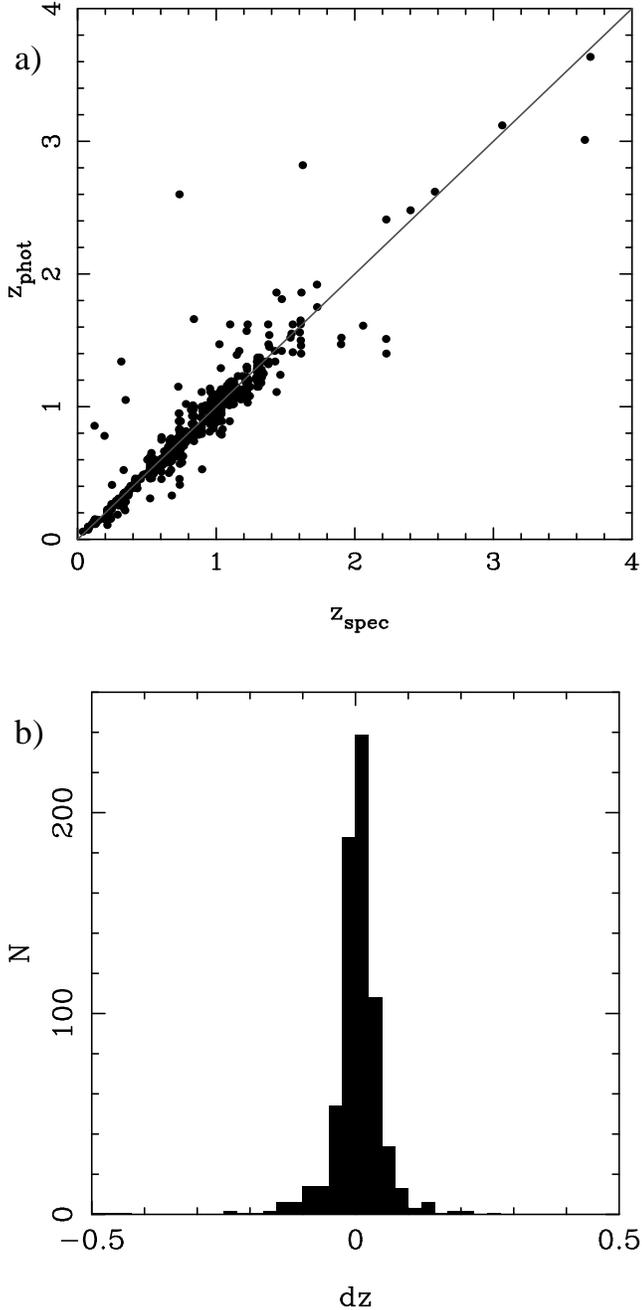}
\caption[]{\label{zqual} a) Photometric versus spectroscopic redshifts for $K_s<21.5$ galaxies in the GOODS/CDFS.  b) Distribution of relative errors. The median of this distribution is $dz=(z_{spec}-z_{phot})/(1+z_{spec})=0.01$ and the rms is $\sigma=0.03$.}
\end{center}  
\end{figure}

\begin{figure}
\begin{center}
\includegraphics[width=1.0\hsize,angle=0] {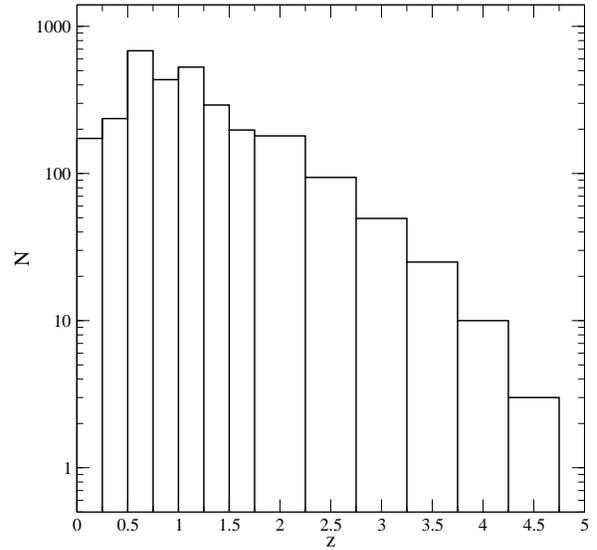}
\caption[]{\label{zhisto} The redshift distribution of $K_s<21.5$ normal galaxies in the GOODS/CDFS.}
\end{center}  
\end{figure}

 We used the public code HYPERZ (Bolzonella, Miralles \& Pell\'{o}, 2000) with the GISSEL98  template library (Bruzual \& Charlot 1993) to model the SED and obtain a photometric  redshift ($z_{phot}$) for each galaxy in our sample, using photometry in up to eight broad-bands ($BVI_{775}z_{850}JK_s$ and IRAC $\rm 3.6 \, \mu m$ and $\rm 4.5 \, \mu m$ bands). Only models with solar metallicity have been considered. Dust-corrections have been taken into account convolving the template SED with the Calzetti et al. (2000) reddening law.

\begin{figure}
\begin{center}
\includegraphics[width=1.0\hsize,angle=0] {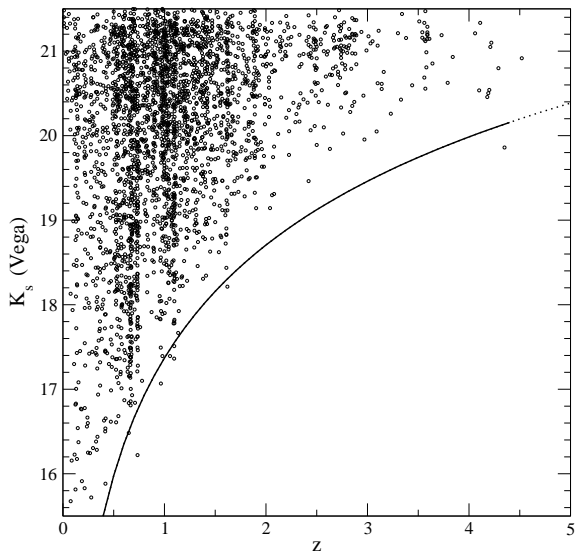}
\caption[]{\label{hubble} Observed $\rm K_s$ magnitudes versus redshifts $z$. The solid line corresponds to the empirical $K-z$ relation for radio galaxies as obtained by Willott et al. (2003), which approximately coincides with the passive evolution of a $\rm 3L^\ast$ starburst formed at redshift $z_f=10$. The dotted line is a nominal extrapolation of the same law.}
\end{center}  
\end{figure}

  As it was shown in Caputi et al.~(2005a), the code HYPERZ has a tendency to favour high-redshift estimates when degenerate solutions in redshift space exist. To control the HYPERZ output, we applied the following set of constraints on the obtained photometric redshifts:

\begin{itemize}

\item We used the $K-z$ relation for radio galaxies (see Figure~\ref{hubble}) to determine  the maximum possible redshift for each object. Radio galaxies delimit the behaviour of the most massive galaxies formed at high redshifts (e.g. Willott et al. 2003) and normal galaxies are not usually found below this limit (Caputi et al. 2004; Caputi et al. 2005a). 

\item The estimated redshift was constrained to the best-fit solution  with a maximum value $z_{phot}=2$ for all those galaxies with a counterpart in the public shallow U-band catalogues existing for the CDFS\footnote{Although $U$-band images  exist for the CDFS (Arnouts et al. 2001), they are shallower and  poorer in resolution than the ACS and ISAAC images used here and, thus, we decided not to incorporate these data for the input catalogue of the photometric redshift algorithms.}, as higher redshift sources are unlikely to be bright at such short wavelengths. Similarly, the estimated redshift was constrained to a maximum value $z_{phot}=4$ for those sources detected in the $B$ band.

\item We used the public code BPZ (Ben\'{\i}tez 2000) to obtain a second, independent,  set of redshift estimates for all the sources in our catalog. In the cases of sources with HYPERZ photometric redshifts $z_{phot}>2$ not confirmed by BPZ, we adopted the lower estimates given by the BPZ code. 

\item We used public COMBO17 photometric redshifts (Wolf et al. 2004) to replace the estimates of those sources with magnitude $R<23.5$ (Vega) and $z_{C17}<1$ (the best regime for COMBO17 redshifts).

\end{itemize}

   Figure~\ref{zqual}a) shows the resulting estimated redshifts versus spectroscopic redshifts for all those $K_s<21.5$ sources included in different publicly available spectroscopy samples in the GOODS/CDFS (e.g. Cimatti et al. 2002b; Le F\`evre et al. 2004; Vanzella et al. 2005). We observe a very good agreement between estimated and real redshifts in most cases.  Figure ~\ref{zqual}b) shows the corresponding distribution of relative errors: the median is $dz=(z_{spec}-z_{phot})/(1+z_{spec})=0.01$ and the standard deviation is $\sigma=0.03$.

   However, in spite of the good accuracy obtained for the photometric redshifts, the incorporation of existing spectroscopic redshifts would be of benefit for the determination of our galaxy redshift distribution. Consequently, in order to optimise the quality of our final redshift catalog, we used the  available spectroscopic redshifts for the CDFS to replace the photometric estimates of our galaxies whenever possible.

   The final redshift catalogue for our 2905 $K_s<21.5$ galaxies is composed of 686/2905 (23.6\%) spectroscopic redshifts, 629/2905 (21.7\%) COMBO17 redshifts and 1590/2905 (54.7\%) HYPERZ/BPZ redshifts.

  Figure \ref{zhisto} shows the resulting redshift distribution of our sample of 2905 $K_s<21.5$ normal galaxies in the GOODS/CDFS. ~81\% of these galaxies lie at redshifts $z \leq 1.5$, while the remaining ~19\% is found to be at $z > 1.5$. The two peaks of the distribution in the redshift bins $[0.5;0.75]$ and $[1.0;1.25]$ are produced by known large-scale structure in the CDFS (e.g. Le F\`evre et al. 2004).  The resulting Hubble diagram $K_s$ versus $z$ is shown in Figure~\ref{hubble}.

\section{The evolution of the $K_s$-band luminosities}
\label{sec_lf}

\begin{table*}
\caption[]{\label{tab_sch} The best-fit parameters for the ML analysis of the rest-frame $K_s$-band LF. The values of $\alpha$ and  $M^\ast(z=0)$  have been taken from Kochanek et al. (2001). $\alpha$ is considered to be constant with redshift. The remaining parameters have been taken as free for the modelling. $\phi_0(z=0)$ is the LF normalisation parameter, for which we obtain a best-fit value in excellent agreement with Kochanek et al. (2001).
}
\begin{tabular}{ccccccc}
\hline
\hline
$\alpha$ & $M^\ast(z=0)$ & $k_\phi$ & $z_\phi$ & $k_M$ & $z_M$ & $\phi_0(z=0)\, \rm (Mpc^{-3})$  \\
\hline
-1.09 (constant) & $-24.16 \pm 0.05$ & $1.67 \pm 0.08$ & $1.78 \pm 0.06$ & $0.63 \pm 0.10$ & $1.88 \pm 0.10$ & $(3.9 \pm 0.3) \times 10^{-3}$ \\
\hline
\hline
\end{tabular}
\end{table*}

   Caputi et al.~(2005a) computed and analysed the rest-frame  $K_s$-band luminosity function (LF) from redshifts $\langle z \rangle =1$ to $\langle z \rangle =2.5$.  In this work, we take advantage of the  larger area covered by the $K_s$-band GOODS/EIS v1.0 release to set tighter constraints on the galaxy rest-frame $K_s$-band LF, and to model its evolution up to $\langle z \rangle =2.5$. The incorporation of IRAC $\rm 3.6 \, \mu m$ and $\rm 4.5 \, \mu m$ data enables us to compute this LF with no need of magnitude extrapolations up to redshifts $z \sim 0.6$ and $z \sim 1$, respectively, and with smaller extrapolations at higher redshifts. Anyway, we note that the k-corrections in the $K_s$-band are less dramatic than those in other shorter wavelength bands (Poggianti et al. 1997).

\begin{figure}
\begin{center}
\includegraphics[width=1.0\hsize,angle=0] {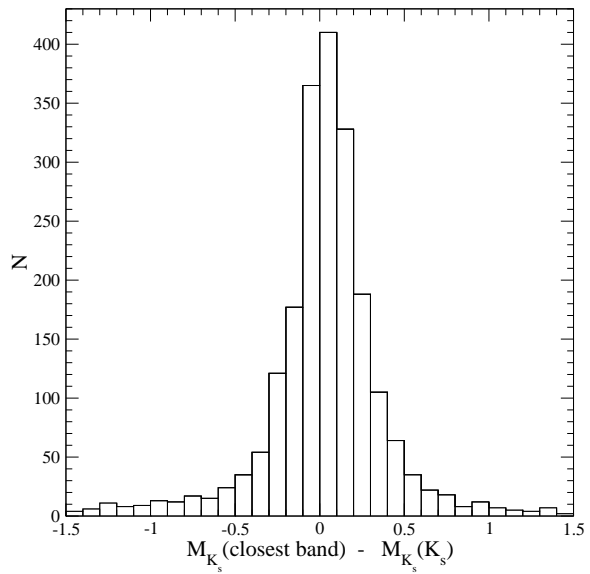}
\caption[]{\label{MKsMKs} The distribution of the differences between the rest-frame absolute $K_s$-band  magnitudes $M_{K_s}$ obtained from the closest available rest-frame band and the ones obtained extrapolating from the observed $K_s$ magnitudes in all cases. The distribution is centered at $M_{K_s}$ (closest) $-M_{K_s} (Ks) = 0.04$ and the rms is $\sigma=0.24$.}
\end{center}  
\end{figure}

    We obtained the rest-frame absolute magnitude $M_{K_s}$ of each galaxy directly from the  HYPERZ output. For those objects with spectroscopic, COMBO17 or BPZ redshifts, we adopted the best SED fitted by HYPERZ constrained to the predetermined redshift.   For the remaining objects, the best-SED fit and redshift estimation were simultaneously obtained. HYPERZ computes the rest-frame absolute magnitude in the selected filter from the observed passband which is the closest to that filter in the rest-frame.  In our case, the rest-frame $M_{K_s}$ magnitudes are obtained either from the $K_s$, $\rm 3.6 \, \mu m$ or $\rm 4.5 \, \mu m$  apparent magnitudes, depending on the redshift of the source and the availability of IRAC data. With this `closest band' method, magnitude extrapolations are minimised.

   It is interesting to compare the  rest-frame absolute $K_s$-band magnitudes $M_{K_s}$ obtained with this procedure, and those obtained when no IRAC data is taken into account for the SED fitting, i.e when the rest-frame $K_s$-band magnitude extrapolations are directly made from the observed $K_s$-band in all cases (as, for instance, in Caputi et al. 2005a, where IRAC data were not available).  Figure~\ref{MKsMKs} shows the distribution of the difference between the absolute magnitudes $M_{K_s}$ obtained with the two different methods. The statistics has been made over the magnitude range relevant for the computation of our LF (see below), i.e. over all the sources with $-27 \leq M_{K_s} \leq -22$, as obtained with the `closest-band' method used here.  We see that, although the resulting distribution has a long tail, the standard deviation $\sigma$ is small: the center of the distribution is located at $M_{K_s}$ (closest) $-M_{K_s} (Ks) = 0.04$ and the rms is $\sigma=0.24$.  Thus, we conclude that, in most cases, the rest-frame absolute magnitudes  $M_{K_s}$  obtained extrapolating  the observed $K_s$-band magnitudes are consistent with those derived from the longer-wavelength data used here.

     We computed the rest-frame $K_s$-band LF using two different techniques: the $1/V_{max}$ method and a maximum likelihood (ML) analysis  (Marshall et al. 1983). In the $1/V_{max}$ method, the LF is directly computed from the data, with no parameter dependence or model assumption. The ML analysis, on the other hand, is a parametric technique which requires an {\em a priori} functional form for the LF and its evolution with redshift. We parametrised the rest-frame $K_s$-band LF with a Schechter function (Schechter 1976):

\begin{eqnarray}
\Phi(L) dL & = & \phi_0(z) \left( \frac{L}{L^\ast(z)}\right)^\alpha exp\left[-\frac{L}{L^\ast(z)}\right] d\left(\frac{L}{L^\ast(z)}\right), \nonumber\\
\frac{L}{L^\ast(z)} & = & 10^{-0.4(M-M^\ast(z))}, 
\end{eqnarray}

\noindent and we assumed that the density and luminosity evolved with redshift as

\begin{eqnarray}
\phi_0(z) & = &  \phi_0(z=0) \times exp\left[-\left(\frac{z}{z_{\phi}}\right)^{k_\phi} \right], \\
M^\ast(z) & = & M^\ast(z=0) - \left(\frac{z}{z_M}\right)^{k_M}.
\end{eqnarray}

\noindent  The parameters $\alpha$  and $M^\ast(z=0)$ have been fixed to the local value: $\alpha=-1.09$ and $M^\ast(z=0)=-24.16 \pm 0.05$ (Kochanek et al.~2001). The obtained best-fit values for the remaining (free) parameters are given in Table~\ref{tab_sch}.

\begin{figure*}
\begin{center}
\includegraphics[width=1.0\hsize,angle=0] {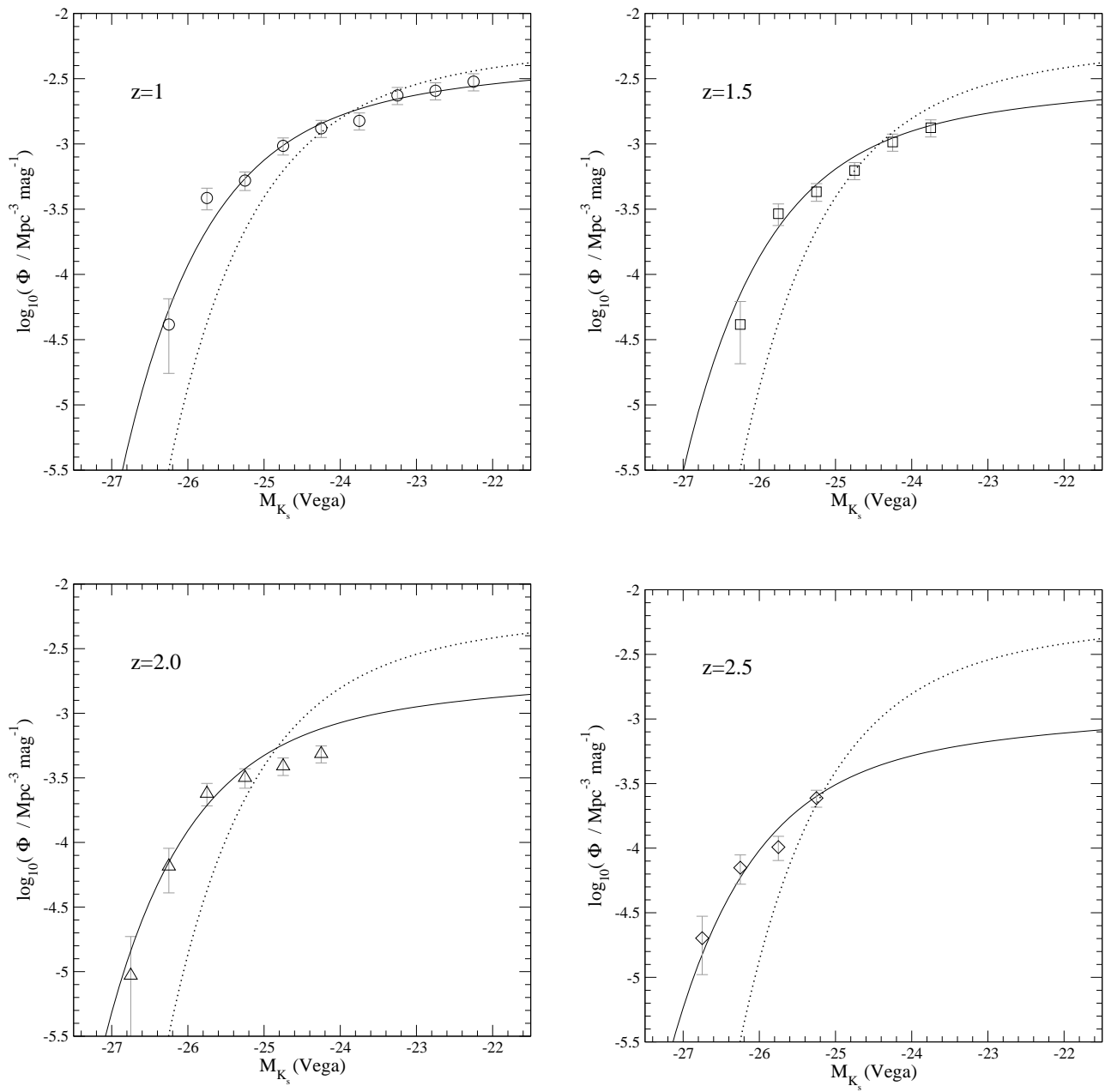}
\caption[]{\label{kslf} The rest-frame $\rm K_s$-band luminosity function at redshifts $\langle z\rangle=1.0, 1.5, 2.0$ and $2.5$ (circles, squares, up triangles and diamonds, respectively).  The empty symbols correspond to the LF computed with the $1/V_{max}$ method. Only values above the luminosity completeness limits of each redshift bin are shown. The solid lines show the resulting LF obtained through  the ML analysis. For a comparison, we show in each panel the local K-band LF measured from 2MASS data by Kochanek et al. (2001; dotted lines).}
\end{center}  
\end{figure*}

 In Figure~\ref{kslf}, we present the resulting rest-frame $K_s$-band LF  at redshifts 
 $\langle z \rangle=1.0,1.5,2.0$ and $2.5$. The empty symbols indicate the LF obtained with the $1/V_{max}$ method. The values shown are strictly above the luminosity completeness limits of each redshift bin.  The number of sources in the luminosity-completeness range are 822, 335, 145 and 83 for the redshift bins $[0.75;1.25], [1.25;1.75], [1.75;2.25]$ and $[2.00;3.00]$, respectively.  To correct for the small incompleteness of our sample to $K_s=21.5$, we weighted each source with a multiplicative factor depending on its apparent magnitude: 1.00, 1.11 and 1.16 for sources with $K_s \leq 20.0$, $20.0 < K_s \leq 21.0$ and $21.0 < K_s \leq 21.5$, respectively (see Section \ref{sec_sample}). These correction factors have a minor impact on the resulting LF (cf. Caputi et al.~2005a). The error bars correspond to the maximum of Poissonian errors and the errors due to cosmic variance. We set a fixed value of 15\% of the number counts for the cosmic variance error bar at $z \gsim 1$ (cf. Somerville et al. 2004). The solid lines show the evolution of the LF with redshift, as obtained through the ML analysis. In the four panels of Figure~\ref{kslf}, we have added for comparison the local K-band LF measured from Two Micron All Sky Survey (2MASS) data by Kochanek et al. (2001; dotted lines).  The LFs obtained by the two different methods are in good agreement in the different redshift bins.

   From the upper-left panel in Figure~\ref{kslf}, we can confirm once more that the bright end ($M_{K_s} \lsim -25$) of the rest-frame $K_s$-band LF significantly increases from redshifts $z=0$ to $z=1$.  This trend has already been observed  in previous works (Drory et al. 2003; Feulner et al. 2003; Pozzetti et al. 2003; Caputi et al. 2004; Caputi et al. 2005a). From our ML analysis we find that, although the total density of objects $\phi_0$ decreases by a factor $\sim 1.5$ between redshifts $z=0$ and $z=1$, the characteristic magnitude $M^\ast$ brightens in $\sim 0.7$ magnitudes between those redshifts. The overall result is a rise in the density of objects with $M_{K_s} \lsim -25$. 
   
   Recently, Dahlen et al. (2005) found that the characteristic magnitude $M^\ast$ of the rest-frame $J$-band LF suffered a dimming from redshifts $z \sim 0.4$ to $z \sim  0.9$. This result was apparently in contradiction with the brightening found by Feulner et al.~(2003) for the same LF in a similar redshift range. 
However, as Dahlen et al. (2005) explain, the one-to-one comparison of the different parameters characterising the Schechter function is misleading, as $M^\ast$, $\alpha$ and $\phi_0$ are coupled, and different behaviours in the evolution of  $M^\ast$  can still produce consistent LFs.   Interestingly, on the other hand, Dahlen et al. (2005) infer that, given the dimming they find for $M^\ast$ in the rest-frame $J$-band LF, the rest-frame $K_s$-band LF should present a similar behaviour.  They suggested that the brightening observed by different authors in the rest-frame $K_s$-band LF might be due to inaccuracies in the k-corrections used to extrapolate  from the observed to the rest-frame $K_s$-band at $z \sim 1$, rather than being a real effect. The incorporation of IRAC $\rm 3.6 \, \mu m$ and $\rm 4.5 \, \mu m$ data in the present work allows us to determine that this is not the case. Once more, one should conclude that a one-to-one comparison of the best-fit parameters obtained  for the Schechter function  (and even more if they are inferred) is misleading. The comparison of the densities of objects in each luminosity and redshift  bin found by different surveys is the correct way of concluding on the resulting LFs.

    As it was shown by Caputi et al. (2004; 2005a), after the rise of the bright end of the rest-frame $K_s$-band LF from redshifts $z=0$ to $\langle z \rangle=1$,  the very bright end ($M_{K_s} \lsim -26$) of this LF does not show any sign of decline up to at least redshift $\langle z \rangle=2.5$. Our ML analysis suggests that the density of the $M_{K_s} \lsim -26$ objects might increase  up to $\langle z \rangle=1.5$ and then stay constant up to $\langle z \rangle=2.5$ within the error bars (cf. Figure~\ref{kslf}). Again, this can be explained by the combined effect of the number and luminosity evolution we obtain through the ML analysis: the total density of objects $\phi_0$ decreases by a factor $\sim 6$ between $z=0$ to $\langle z \rangle=2.5$, but the  characteristic magnitude $M^\ast(z)$  brightens by more than  one magnitude between these redshifts.

\begin{figure*}
\begin{center}
\includegraphics[width=1.0\hsize,angle=0] {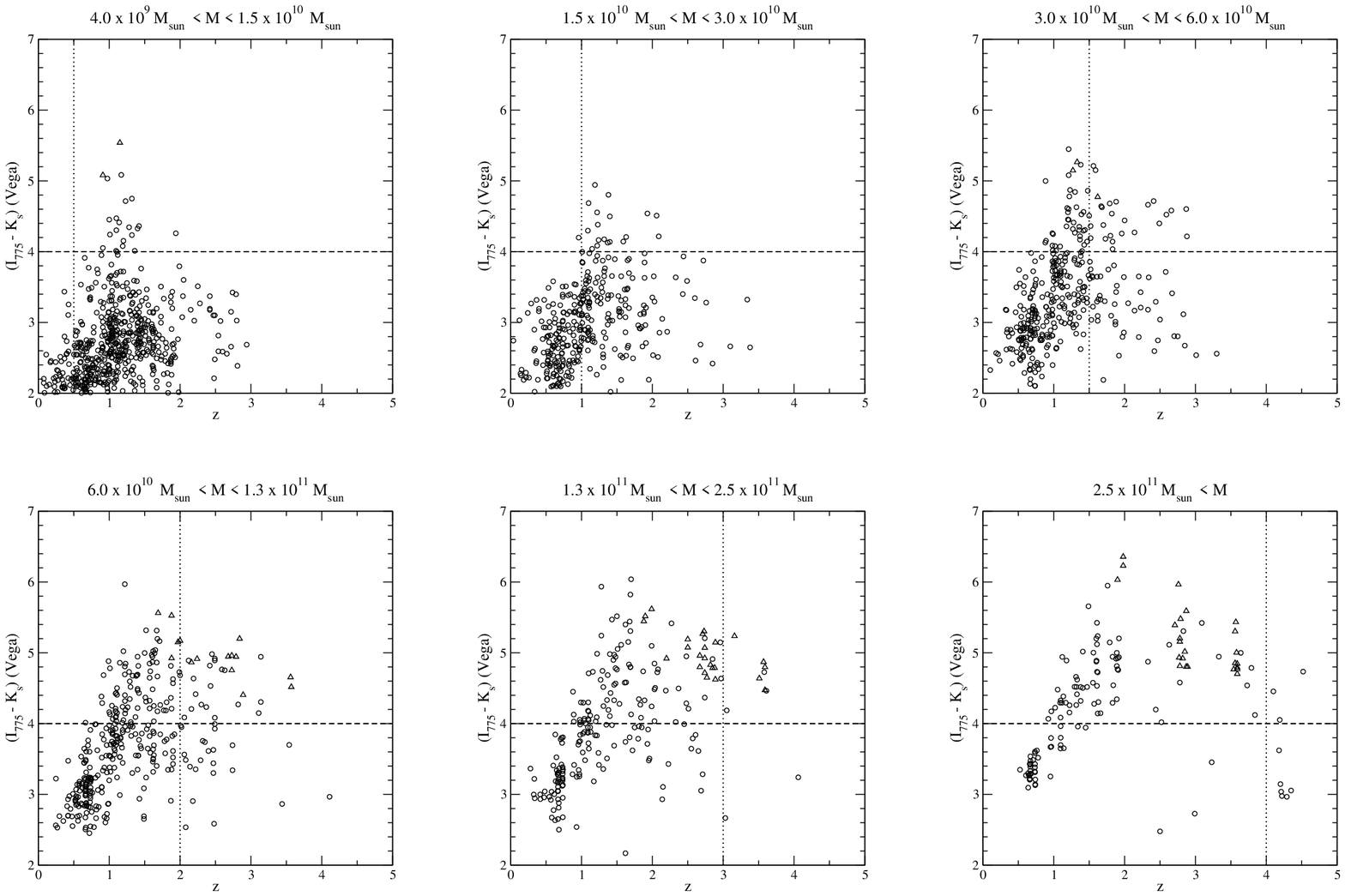}
\caption[]{\label{imKmass} The observed $(I_{775}-K_s)$ colours versus redshift for galaxies within different ranges of stellar mass.  In all the panels, circles and triangles indicate precise and lower limits to the $(I_{775}-K_s)$ colours, respectively. The horizontal dashed line represents the typical extremely-red-galaxy colour cut, $(I_{775}-K_s)=4$, while the vertical dotted line delimits the region of completeness for each stellar mass bin.}
\end{center}  
\end{figure*}

   The combined density and luminosity evolution we determined through the ML analysis also implies a particularly strong decrement in the number of intermediate and low luminosity ($M_{K_s}>-25$) objects with increasing redshift.  We remark, however, that  we do not directly see these  intermediate and low luminosity objects at high redshifts, given the limits of our survey. We only extrapolate this behaviour from the LF  obtained through the ML analysis, which assumes the validity of a Schechter function for the description of the LF at all redshifts. Thus, two alternative conclusions can be extracted from this result: 1) a significant fraction of low and intermediate luminosity objects are created between $\langle z \rangle=2.5$ and $z=0$, or 2) at least as many low luminosity objects are present at high as at low redshifts, and our assumptions for the computation of the LF with the ML analysis are not valid to describe the faint-end  of the LF at high redshifts. The direct access to low luminosity objects at high redshifts through deeper surveys is necessary to conclusively prove the behaviour of the faint-end of the LF.  
   
    In hierarchical models, the creation of new  objects is the consequence of the production of galaxy mergers and subsequent triggering of star formation. Very  recently, Conselice (2005) found that the fraction of mergers in faint and low mass galaxies peaked at lower redshifts  than for higher luminosity and more massive galaxies. This result appears to be consistent with the presence of luminous galaxies at high redshifts $z \sim 2.5$, as they would be produced from mergers at these or even higher redshifts. The decline in the number of intermediate luminosity objects at high redshifts that we extrapolate from the ML analysis would also be consistent with the fact that a significant fraction of them form through mergers at later times. However, the differential redshift peak observed for mergers of different mass and luminosity still cannot decide on the formation epoch of all the lowest luminosity objects, and the two hypotheses discussed above could still be possible. At redshifts $z \lsim 1$, major mergers become progressively less important (e.g. Patton et al. 2002), so it is expected that the evolution of the $K_s$-band LF is mainly governed by the dimming of the already present old stellar populations. The significant decrease of the bright end ($M_{K_s}<-25$) of the LF from redshifts $\langle z \rangle=1$ to $z=0$ is probably indicating that the most massive galaxies present at $z\sim 1$ must have finished the construction of the bulk of their stars by this redshift, and mainly fade out all the way through redshift $z=0$ (cf. Section~\ref{sec_masscol}).

\section{The evolution of the stellar mass}
\label{sec_mass}

\subsection{Computation of stellar masses}

 We used the rest-frame $K_s$-band luminosity of each galaxy to compute its stellar mass.  We obtained the rest-frame absolute magnitude $M_{K_s}$ (and corresponding luminosity) from the HYPERZ best-fit SED on an individual basis, as explained in Section \ref{sec_lf}. We used the public code GALAXEV (Bruzual \& Charlot 2003) to construct grids of mass-to-light ($M/L_{K_s}$) ratios depending on galaxy age, assuming a Salpeter IMF over stellar masses $M=(0.1-100) \, M_\odot$. The advantage of using luminosities in the rest-fame $K_s$-band is that other parameters, such as the dust-correction or the galaxy star-formation history, have a minor importance in the determination of the $M/L_{K_s}$ ratios.  As in Caputi et al.~(2005a), we modelled the evolution of these ratios using two galaxy templates, with no dust, corresponding to different exponentially declining star-formation histories with characteristic times $\rm \tau \sim 0.1 \, Gyr$ and $\rm \tau \sim 5 \, Gyr$. We determined the $M/L_{K_s}$ of  each galaxy using the model most similar to the HYPERZ best-fit template in each case, and interpolating between the values which corresponded  to the closest ages in the grids. Our estimated stellar masses  are typically accurate within a factor $\lsim 2$ (where the larger uncertainties correspond to plausible degeneracies in parameter space, especially in those cases in which the galaxy redshift is simultaneously determined with the best-SED fit). On the contrary, the dependence on star-formation history and dust corrections would be much stronger if optical rather than near-IR luminosities were used to derive the stellar masses.

 As we discussed in Section~\ref{sec_lf},  the rest-frame $K_s$-band luminosity derived as here and the luminosity  that would be derived from the extrapolation of the  corresponding observed $K_s$-band magnitude are consistent within a rms $\sigma=0.24$. This implies that the derived stellar masses  will be consistent within $10^{-0.4 \times (\pm 0.24)}$, i.e. within a factor $\sim 1.25$. Thus, we conclude that the use of the observed B through $K_s$-band data, with no incorporation of IRAC data, would have introduced an additional uncertainty of $\sim 25$\% on the derived stellar masses. This is  a  relatively minor contribution to the error budget, which is mainly dominated by plausible degeneracies in parameter space and especially by the adoption of a fixed IMF.

\subsection{The colours and stellar masses of galaxies at different redshifts}
\label{sec_masscol}

  Figure~\ref{imKmass} shows the observed ($I_{775}-K_s$) colours versus redshift for galaxies with stellar masses within different mass ranges. In all the panels, circles represent those objects with precise ($I_{775}-K_s$) colours, while triangles indicate lower limits to the colours of objects with magnitude $I_{775}>26$ (Vega; equivalent to $\lsim 2\sigma$ detections in the $I_{775}$ band), computed as ($26-K_s$).  The horizontal dashed line shows the typical colour cut defining extremely red galaxies (ERGs; $(I_{775}-K_s)>4$, Vega), while the vertical dotted lines delimit the region of completeness for the lowest masses in each mass bin. The mass completeness limits  have been estimated using the median of the k-corrections and the maximal  $M/L_{K_s}$ ratios at each redshift (corresponding to a galaxy formed in a single burst at redshift $z_f=\infty$).   At redshifts $z \gsim 1$, typical ERG colours  indicate the presence of old stellar populations and/or young stellar populations heavily enshrouded by dust. On the contrary, $(I_{775}-K_s)<4$ colours at $z>1$ imply the presence of unextincted on-going star-formation. At $z \lsim 1$, the observed $(I_{775}-K_s)=4$ colour cut is merely formal, as only systems with very large dust extinctions would show colours above that limit.

  Inspection  of Figure~\ref{imKmass} shows that the $(I_{775}-K_s)$ colours of galaxies at a given redshift depend on the stellar mass. The fraction of red-to-blue galaxies systematically increases with assembled stellar mass, even considering each mass bin only up to the completeness redshift limit. In particular, the vast majority of galaxies with $M>2.5 \times 10^{11} \, M_\odot$  between redshifts $z=1$ and $z=4$ are ERGs. Some of these galaxies might be purely passive systems (cf. e.g. Daddi et al. 2005), while in other cases the extremely red colours would only be explained by the superposition of old stellar components and dust (see Caputi et al. 2004; Papovich et al. 2005; and cf. Section~\ref{sec_nearir}).  The extremely red colours characterising these  very massive galaxies reveal important constraints on their star-formation histories.   In  contrast to what is observed in other less massive systems, our results indicate that unobscured time-extended star-formation at $z \gsim 1$ is rare in galaxies with assembled stellar mass $M>2.5 \times 10^{11} \, M_\odot$. The main build-up of these systems must be produced in short bursts and/or with considerable amounts of dust.

\begin{figure}
\begin{center}
\includegraphics[width=1.0\hsize,angle=0] {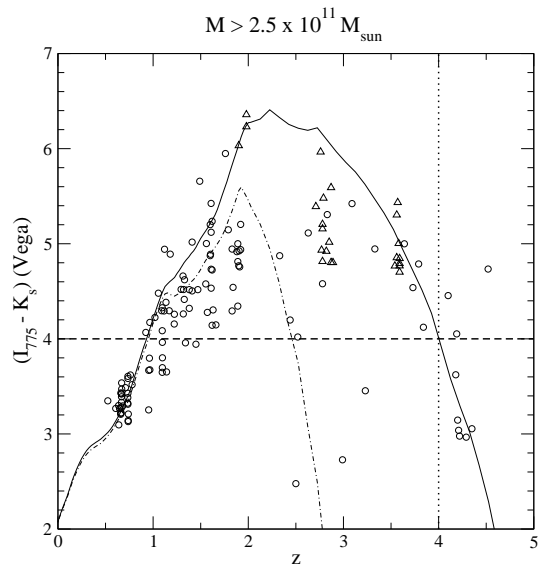}
\caption[]{\label{imKhighm} The observed $(I_{775}-K_s)$ colours versus redshift for galaxies with estimated stellar mass $M>2.5 \times 10^{11}\, M_\odot$. The symbols are the same as in Fig.~\ref{imKmass}. The solid and dotted-dashed lines show the modelled evolution of the $(I_{775}-K_s)$ colours for galaxies formed in an instantaneous burst at redshift $z_f=5$ and $z_f=3$, respectively, both with no dust and passive evolution thereafter.}
\end{center}  
\end{figure}

  Figure~\ref{imKhighm} shows an amplified version of the lower-right panel of Figure~\ref{imKmass}. For reference, we added in this figure the modelled  evolution of the $(I_{775}-K_s)$ colours for galaxies formed in an instantaneous burst at redshift $z_f=5$ and $z_f=3$ (solid and dotted-dashed lines, respectively), both with no dust and passive evolution thereafter. From Figure~\ref{imKhighm} we can see that, at $z>1$, galaxies within the ERG regime with assembled stellar masses $M>2.5 \times 10^{11}\, M_\odot$ display a wide range of $(I_{775}-K_s)$  colours. This suggests that they must undergo a variety of  star-formation histories, either different formation redshifts or at least different formation redshifts for the youngest stellar population present in them.

   At redshifts $z \lsim 1$, on the contrary, the galaxies with stellar mass  $M>2.5 \times 10^{11}\, M_\odot$ present in our sample show quite homogeneous $(I_{775}-K_s)$ colours. This is again in contrast to what is observed in less massive systems. The homogeneous colours displayed by these most massive galaxies suggest that their optical/near-IR light must be dominated by old stellar populations.  However, this does not  necessarily  mean that all of them are following a strictly passive evolution, but that any possible on-going star formation would  only 
   produce minor changes to their properties (e.g. stellar mass). In fact, a substantial fraction  ($ \gsim 30\%$) of these galaxies are detected in the mid-IR, i.e. experience on-going star-formation (Caputi et al. 2005b, 2005c). The typical IR-derived star-formation rates of these  mid-IR detected galaxies at $z<1$ are $10-50 \, M_\odot \, \rm yr^{-1}$ (Caputi et al. 2005b). If we consider that the lifetime of a starburst episode is at most $\sim 10^8 \, \rm yr$, we infer that these galaxies should be forming  $1-5 \, \times 10^9 \, M_\odot$ per starburst episode at those redshifts, i.e. an  additional minor amount to their already existing stellar mass. Some of these  very massive systems at $z \lsim 1$ could be candidates for cluster E/S0 galaxies, as early-type cluster galaxies show very little dispersion in their colours from $z=0$ (e.g. Bower et al. 1992) to  $z \sim 1$ (e.g. Ellis et al. 1997; Stanford et al. 1998; Blakeslee et al. 2003).  Indeed, the existence of large-scale structure is suggested by the colour-redshift diagram, as a considerable number of our $M>2.5 \times 10^{11} \, M_\odot$ galaxies appear to lie at discrete redshift values. On the contrary, only a morphology study would be able to conclude on the plausible early or late-type classification of these systems (McLure et al. 2005, in preparation).

\subsection{The evolution of the galaxy number and mass densities}   
   
\begin{figure}
\begin{center}
\includegraphics[width=1.0\hsize,angle=0] {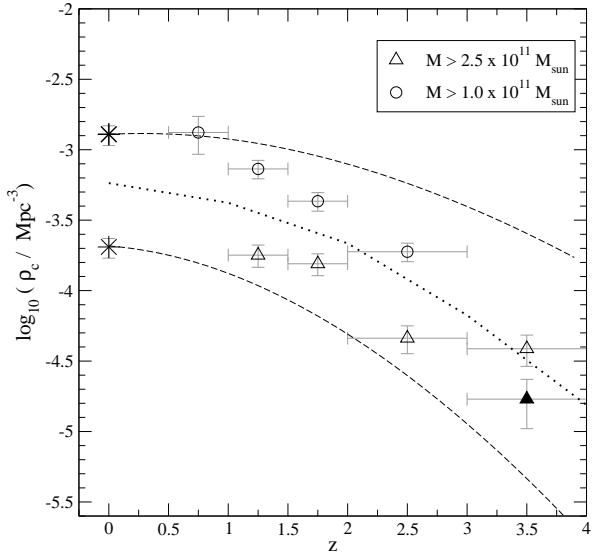}
\caption[]{\label{cnumd} The comoving number density of galaxies with stellar mass $\rm M>1 \times 10^{11} M_\odot$ (circles) and $\rm M>2.5 \times 10^{11} M_\odot$ (triangles) (Salpeter 1955 IMF, restricted to $M=(0.1-100)\, M_\odot$). The filled triangle at $z=3.5$ denotes the re-computed number density of  $\rm M>2.5 \times 10^{11} M_\odot$ galaxies after correcting for the effects of large-scale structure in the corresponding redshift bin. The star-like symbols  indicate the density values for the same mass thresholds at $z=0$, obtained integrating the local stellar mass function (Cole et al. 2001; Bell et al. 2003).   The dashed lines show the evolution of the number density of haloes with total mass $\rm M>2 \times 10^{12} M_\odot$ and $\rm M>1 \times 10^{13} M_\odot$ (upper and lower curves, respectively), as obtained from  $\Lambda$-CDM models of structure formation (with a code kindly provided by Will Percival). The dotted line corresponds to the predicted number density of galaxies with mass $M>1 \times 10^{11} M_\odot$ obtained with SPH simulations (Nagamine et al. 2005).}
\end{center}  
\end{figure}

\begin{figure}
\begin{center}
\includegraphics[width=1.0\hsize,angle=0] {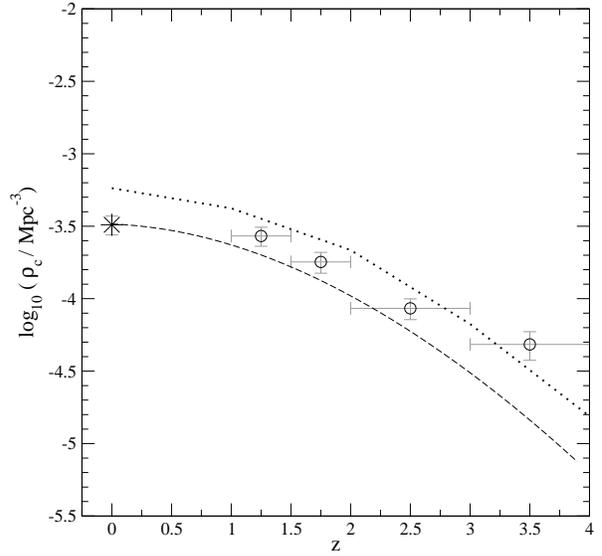}
\caption[]{\label{chabrier} The comoving number density of galaxies with stellar mass $\rm M>1 \times 10^{11} M_\odot$ (Chabrier 2003 IMF).   The symbols are the same as in Figure~\ref{cnumd}. 
}
\end{center}  
\end{figure}

   As we did in Caputi et al.~(2005a), we analyse the evolution of the number density of different-mass galaxies over cosmic time. In the present work,  however, due to the analysis of a considerably larger area,  we are able to select a statistically significant sample of very massive galaxies ($\rm M>2.5 \times 10^{11} M_\odot$), which enables us to explore the very-high mass end of the  stellar-mass evolution.

    Figure~\ref{cnumd} shows the comoving number density of galaxies with assembled stellar mass $M>1 \times 10^{11} M_\odot$ (circles) and $\rm M>2.5 \times 10^{11} M_\odot$ (empty triangles). The asterisks denote the corresponding local values, as obtained from the integration of the local stellar mass function (Cole et al. 2001; Bell et al. 2003). As in the computation of the galaxy LF (cf. Section \ref{sec_lf}), the error bars correspond to the maximum of Poissonian errors and the errors due to cosmic variance.  All the results presented in this section incorporate $V_{max}$ corrections, but deliberately not sample incompleteness corrections (which anyway are very small).  We decided to do so to explicitly show that the contrast betweeen some of the observational and theoretical results discussed below is independent of such corrections.
   
   The evolution of the densities of galaxies with stellar mass $M>1 \times 10^{11} M_\odot$ shows that a minimum of 15\% of the local value is in place before redshift $z=3$ (this value rises to $\sim 20$\% considering the number densities cut at stellar masses $M>1.3 \times 10^{11} M_\odot$, i.e. the regime of strict mass completeness to $z=3$). This results is consistent with Caputi et al.~(2005a). However, in contrast to this previous work, we observe here that the number density of  $M>1 \times 10^{11} M_\odot$ systems does increase from redshifts $z \approx 3$ to   $z \approx 1.5$, although this increment is modest  ($\sim 20$\% of the local value). More than 50\% of the local $M>1 \times 10^{11} M_\odot$ galaxies appear to have formed at redshifts $z \lsim 1.5$.

   The dashed lines in Figure~\ref{cnumd}  trace the evolution of dark-matter haloes with total mass $\rm M>2 \times 10^{12} M_\odot$  and $\rm M>1 \times 10^{13} M_\odot$ (upper and lower curves, respectively), as obtained from the extended Press-Schechter formalism (Bond et al.~1991), with a code kindly provided  to us by Will Percival. The mass thresholds for the halo distributions have been deliberately selected to coincide with the corresponding number densities of galaxies at $z=0$, assuming a single halo occupation number.  The compared observed and predicted density evolutions  confirm the trend found in Caputi et al.~(2005a): at redshifts $1 \lsim z \lsim 3$, the density of haloes with total mass $\rm M>2 \times 10^{12} M_\odot$ is systematically larger than the density of galaxies with stellar mass $M>1 \times 10^{11} M_\odot$. Caputi et al. (2005a) also showed that the number density of galaxies and haloes coincide again at higher redshifts $z \sim 3.5$, although this cannot be re-assessed from the present sample as the  $K_s=21.5$ cut implies completeness only for stellar masses $\rm M>2.5 \times 10^{11} M_\odot$ up to $z \approx 4$.

     On the other hand, we investigate the very high mass end ($\rm M>2.5 \times 10^{11} M_\odot$) of galaxy number density evolution and also compare with the corresponding theoretical predictions. Inspection of Figure~\ref{cnumd} shows that up to $\sim 20$\% of the local massive galaxies with assembled stellar mass $\rm M>2.5 \times 10^{11} M_\odot$    is in place before $z=4$, while the construction of the majority of these very massive systems occurs between redshifts $z=3$ and $z=1.5$. Within our sample, we find that $\sim 90$\% of the local galaxies with mass $\rm M>2.5 \times 10^{11} M_\odot$ is already in place by redshift $z \approx 1$.  This is contrast to what is observed for less massive systems, whose evolution is only halfway complete by that redshift.

    The comparison with theoretical models shows that, if we assume the same conversion between baryonic and dark-matter halo mass as inferred at $z=0$, there appear to be too few
dark matter haloes of adequate mass in place at high redshift to host the
observed number of high mass galaxies. We observe that the deficit is particularly important at $z \gsim 3$. Similar incompatibilites between the presence of very massive systems   and the predictions of theoretical models at high redshifts have already been pointed out in previous studies  (e.g. Fontana et al.~2003; Genzel et al.~2003; Pozzetti et al.~2003; Cimatti et al.~2004; Fontana et al.~2004; McCarthy et al.~2004; Saracco et al.~2005). Our results show that this incompatibility might certainly exist for  the most massive galaxies. It could be argued, however,  that our excess of very massive galaxies might be biased by cosmic variance effects (more than what is accounted for within the error bars). In effect, as we noted in Section~\ref{sec_masscol}, the existence of large-scale structure within our sample  can be deduced from the colour-redshift diagrams.  Within our sample, we find 16 galaxies with $\rm M>2.5 \times 10^{11} M_\odot$ at $3<z<4$, yielding a number density $\rho_c=(3.87 \pm 0.97)\times 10^{-5}\, {\rm Mpc^{-3}}$. 9/16 of these galaxies have photometric redshifts concentrated in the narrow interval  $3.55<z<3.65$, and there is also a spectroscopically confirmed QSO at redshift $z=3.61$ within our surveyed field (not included in the final galaxy sample analysed in this work), strongly suggesting  the presence  of large-scale structure for very massive galaxies at these redshifts.  To obtain a lower limit on the number density  of galaxies with stellar mass $\rm M>2.5 \times 10^{11} M_\odot$ at $\langle z \rangle=3.5$, we re-computed this density excluding the 9 galaxies at $3.55<z<3.65$. We obtained $\rho_c=(1.69 \pm 0.64)\times 10^{-5}\, {\rm Mpc^{-3}}$, i.e. $\log_{10}(\rho_c)=-4.77^{+0.14}_{-0.21}$ (filled triangle in Figure~\ref{cnumd}), while the density of the corresponding haloes at that redshift is only $\log_{10}(\rho_c)=-5.33$. Thus, it seems that even taking into account the effects of large-scale structure, we cannot completely reconcile the theory predictions with the observations at high redshifts. 
 
\begin{figure}
\begin{center}
\includegraphics[width=1.0\hsize,angle=0] {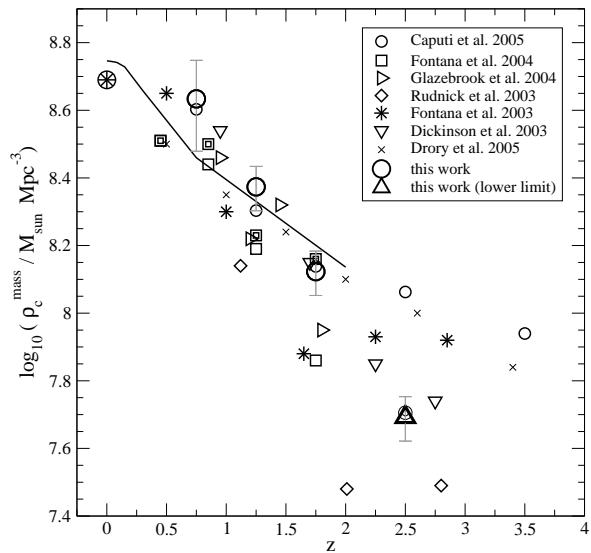}
\caption[]{\label{compstm}  The  total stellar mass densities obtained from the present study (large circles and up-triangle, with error bars, for precise values and lower limit, respectively), compared to values previously obtained by other authors: Dickinson et al. 2003 (down-triangles); Fontana et al. 2003 (asterisks); Rudnick et al. 2003 (diamonds); Glazebrook et al. 2004 (right-triangles); Fontana et al. 2004 (squares and double squares for the observed and extrapolated values in the K20 survey, respectively); Caputi et al. 2005a (small circles and double circle for total and $M>10^{11}\, M_\odot$ mass densities, respectively); Drory et al. 2005 (crosses).  The circle with a star at $z_{phot}=0$ show the local value of the total stellar mass density obtained from the integration of the local galaxy stellar mass function (Cole et al. 2001, Bell et al. 2003). The solid line is the evolution of the stellar mass density as obtained from the integration of  the star-formation rates derived from Sloan Digital Sky Survey (SDSS) data (Heavens et al. 2004). Dickinson et al. values correspond to the average of the values obtained with the 1-component and 2-component models with solar metallicity (see Table 3 in Dickinson et al. 2003). Rudnick et al. values are computed only on galaxies with rest-frame V-band luminosity $\rm L_V > 1.4 \times 10^{10} L_{\odot}$ and, thus, miss a significant fraction of the mass. Glazebrook et al. values only correspond to galaxies with estimated mass $\rm \log(M)>10.46$ and are at least partially incomplete above redshift $z \sim 1.2$.  All the values in this figure assume a Salpeter IMF over stellar masses $M=(0.1-100)\, M_\odot$ and a cosmology with $\rm H_o=70 \,\rm km \, s^{-1} Mpc^{-1}$, $\rm \Omega_M=0.3$ and $\rm \Omega_\Lambda=0.7$.}
\end{center}  
\end{figure}

  Recently, Nagamine et al.~(2005) showed that the properties of high redshift red and massive galaxies could be well reproduced through the use of hydrodynamical simulations with star-formation rate histories that peak at higher redshifts than those used in semianalytic models. It is then interesting to compare also their predictions with our observed evolution of number densities of massive galaxies. The dotted line in Figure~\ref{cnumd} shows the predicted  number density of galaxies with stellar mass $M>1 \times 10^{11} M_\odot$, as obtained with the Nagamine  et al. smoothed particle hydrodynamics (SPH) simulations.  We observe that these simulations yield densities below the observed values, and the discrepancies are larger at low than at high redshifts. 
  
  However, in the Nagamine et al. models, galaxies masses are directly obtained from the simulations and are IMF independent. Thus, for a proper comparison with these model results, we need to test whether the adoption of a different IMF to compute our stellar masses could improve the agreement between model and data.  Figure~\ref{chabrier} shows the evolution of the comoving number density of galaxies with stellar mass  $M>1 \times 10^{11} M_\odot$ considering a Chabrier~(2003) IMF. Both our observed densities and the local value have been transformed in accordance to this IMF. The change from a Salpeter~(1955) to a Chabrier~(2003) IMF produces an overall re-scaling of all our stellar masses by $\sim 0.3 \rm \, dex$, in such a way that a cut $M>1 \times 10^{11} M_\odot$ for masses in the Chabrier IMF corresponds to a cut $M \gsim 2 \times 10^{11} M_\odot$ in the Salpeter IMF.

  We repeat the same exercise as in Figure~\ref{cnumd} and compare the observed values with the densities of dark matter haloes with a mass threshold deliberately selected to coincide with the density of galaxies at $z=0$. As in Figure~\ref{cnumd}, we observe that, at very high redshifts $z \gsim 3$, the predicted number of haloes appears to be insufficient to encompass the observed densities of $M>1 \times 10^{11} M_\odot$ (Chabrier) galaxies. On the contrary, the SPH predicted number densities of $M>1 \times 10^{11} M_\odot$ galaxies  by Nagamine et al. (2005) appear to be in quite good agreement with our observed values from redshifts $z \sim 1$ through $z \sim 4$, although they produce some excess of massive local galaxies. We note that a similar behaviour would be deduced from the densities of dark matter haloes if we did not constrain the comparison to those haloes with densities coinciding with the local densities of massive galaxies. This shows that what we find here is a direct discrepancy with the predictions of hierarchical models, and does not depend on the specific recipes adopted either by semianalytic models or SPH simulations.

  In summary,  hierarchical models do not seem to be able to reproduce the observed densities of the most massive galaxies from the local Universe to very high redshifts. Alternatively, it could also be argued that the use of a single IMF is not suitable to convert from luminosities to stellar masses in the whole range of redshifts $z=0$ to $z=4$ (see e.g. Kroupa \& Weidner 2005, for a discussion). Unluckily, the validity of a single IMF is an issue extremely hard to test at high redshifts, and it is not possible to overcome the IMF uncertainty when doing the comparisons between observations and theory.

  Finally, we computed the evolution of the total stellar mass density with redshift, as derived from our sample of $K_s<21.5$ galaxies. Figure~\ref{compstm} shows our obtained stellar mass densities (large circles and up-triangle, with error bars), compared to other values previously published in the literature.  From the present sample we find that the total stellar mass densities at $\langle z \rangle=1.25$ and $\langle z \rangle=1.75$ are $\rho_c=(2.34 \pm 0.35) \times 10^{8} \, M_\odot \, {\rm Mpc^{-3}}$ and $\rho_c=(1.33 \pm 0.20) \times 10^{8} \, M_\odot \, {\rm Mpc^{-3}}$, respectively, i.e. $(48 \pm 7)\%$ and $(27 \pm 4)\%$ of the local value. These values  very well agree  with the previous determinations made by Caputi et al.~(2005a) and other works.  At higher redshifts, the completeness limits of our sample only allows us to obtain lower limits on the total stellar mass densities. The up-triangle  in Figure~\ref{compstm} represents a lower limit to the total stellar mass density at redshift $\langle z \rangle=2.5$. As our galaxy sample is only complete for masses $M>1.3\times 10^{11} \, M_\odot$ up to $z \approx 3$, we compare this lower limit to the stellar mass density contained in galaxies with stellar mass $M>1\times 10^{11}\, M_\odot$ obtained by Caputi et al. (2005a) (double circle in Figure~\ref{compstm}),  and observe a very good agreement. The total stellar mass densities at redshifts $z \gsim 2$ have been better constrained by previous deeper surveys.  The estimated stellar mass densities at $z \sim 2.5$ and $z \sim 3.5$ are (20-24)\% and (13-18)\% of the local values, respectively (Caputi et al. 2005a; Drory et al. 2005).

\section{Additional hints from pure near-IR colours}
\label{sec_nearir}

\begin{figure}
\begin{center}
\includegraphics[width=1.0\hsize,angle=0] {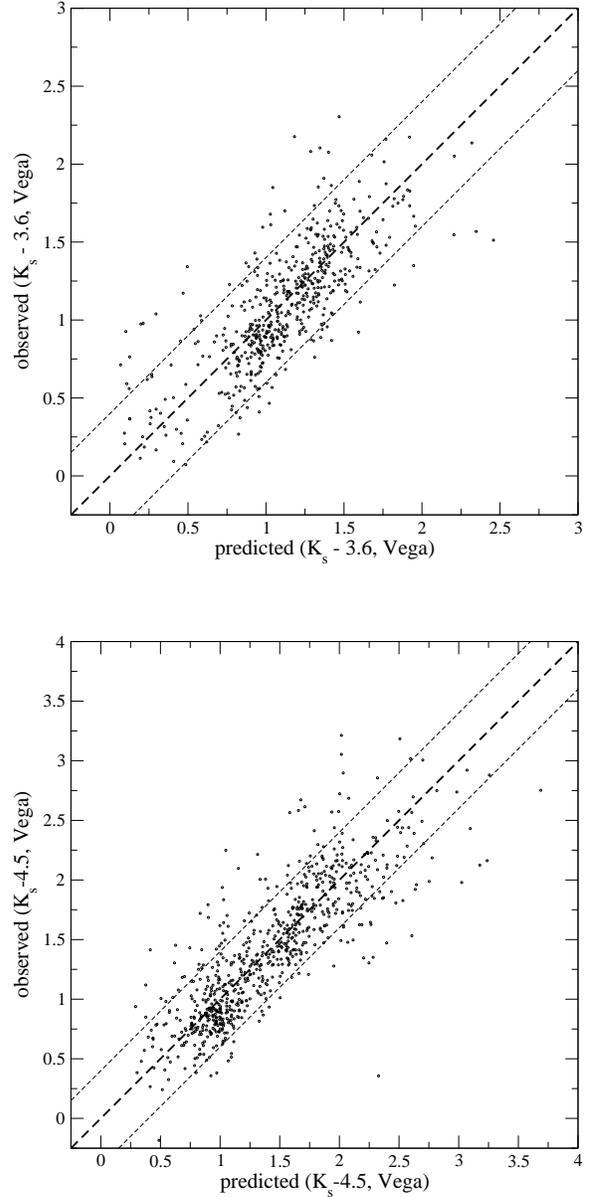}
\caption[]{\label{colobspr} The observed versus predicted $(K_s-[3.6 \, \mu {\rm m}])$ and $(K_s-[4.5 \, \mu {\rm m}])$ colours (upper and lower panels, respectively) for the $K_s<21.5$ (Vega) galaxies in the restricted area of the GOODS/CDFS studied by Caputi et al. (2005a). The dashed lines indicate the region of consistency, taking into account the error bars on the observed colours.}
\end{center}  
\end{figure}

  Caputi et al.~(2005a) predicted the pure near-IR $K_s$/IRAC colours of $K_s$-selected galaxies based on the best-fit SED of each object. These predictions were made for the galaxies selected in a sub-region of 50.4 arcmin$^2$ of the area analysed in this work. In the light of the currently available IRAC data, we can assess the validity of these predictions. Figure~\ref{colobspr} shows the observed versus predicted $(K_s-[3.6 \, \mu {\rm m}])$  and $(K_s-[4.5 \, \mu {\rm m}])$ colours for the $K_s$-band galaxies in the restricted GOODS/CDFS area (upper and lower panels, respectively). For clarity, we only show those sources with small error bars in their observed IRAC magnitudes, i.e. $\varepsilon(3.6 \, \mu {\rm m})<0.3$ and $\varepsilon(4.5 \, \mu {\rm m})<0.3$. As in all cases the $K_s$ magnitudes have $\varepsilon(K_s) \lsim 0.2$, the  measured colours are typically precise within $\sim$ 0.4 mag (dashed lines in both panels of Figure~\ref{colobspr}). From this figure we determine that 89\% and 84\% of the predicted $(K_s-[3.6 \, \mu {\rm m}])$  and $(K_s-[4.5 \, \mu {\rm m}])$ colours, respectively,  are in agreement with the observed colours, within the error bars. We can confirm, then, the majority of the near-IR colours predicted in Caputi et al.~(2005a).

\begin{figure}
\begin{center}
\includegraphics[width=1.0\hsize,angle=0] {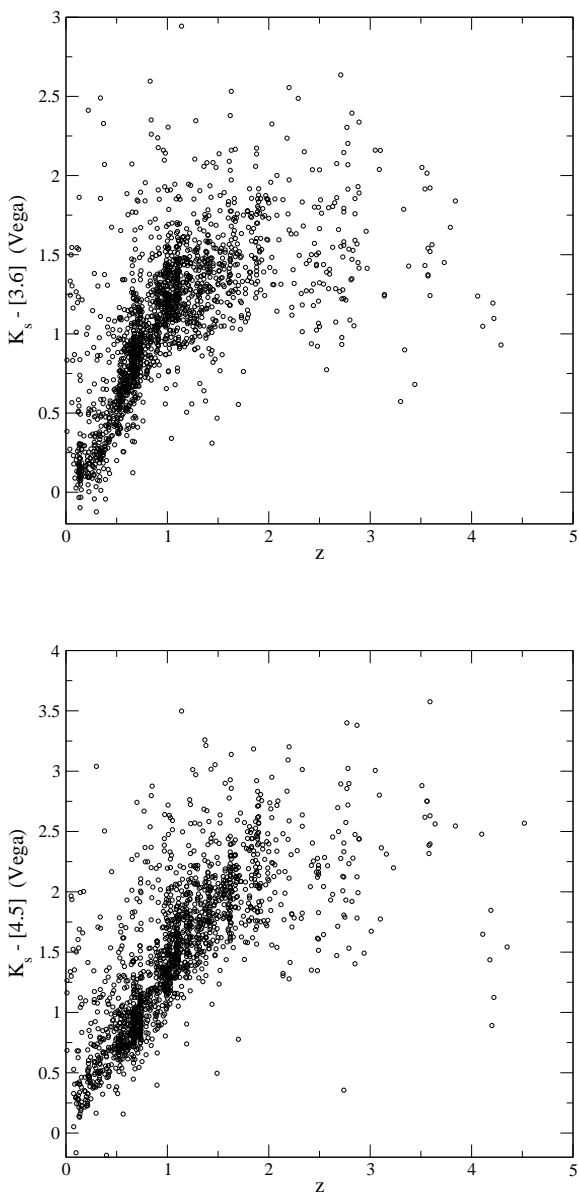}
\caption[]{\label{colnearir} The observed $(K_s-[3.6 \, \mu {\rm m}])$ and $(K_s-[4.5 \, \mu {\rm m}])$ colours (upper and lower panels, respectively)  versus redshift for the $K_s<21.5$ (Vega) galaxies in the GOODS/CDFS.}
\end{center}  
\end{figure}

   Figure~\ref{colnearir} shows the observed $(K_s-[3.6 \, \mu {\rm m}])$  and $(K_s-[4.5 \, \mu {\rm m}])$ colours versus redshift (upper and lower panels, respectively) for the $K_s<21.5$ (Vega) galaxies in the larger sample studied in this work.  $\rm 3.6 \, \mu m$ and $\rm 4.5 \, \mu m$ data are available for alternative $\sim 2/3$ of the sources in our sample. Once more, for clarity, we only show those sources with magnitude  errors $\varepsilon(3.6 \, \mu {\rm m})<0.3$ and $\varepsilon(4.5 \, \mu {\rm m})<0.3$.  The variation of these observed colours with redshift confirms the trends predicted by Caputi et al.~(2005a) within the error bars. Inspection of Figure~\ref{colnearir}  reveals, however,   the existence of a population of low-redshift ($z \lsim 0.5$) galaxies with colours $(K_s-[3.6 \, \mu {\rm m}])>1.0$ and $(K_s-[4.5 \, \mu {\rm m}])>1.5$, which are quite redder than those predicted for low-redshift galaxies in the previous work.    The colour excess of these galaxies cannot be reconciled with the predictions by simply taking into account the photometry error bars. The analysis of these sources shows that  31 out of the 32 objects satisfying $z \lsim 0.5$ and $(K_s-[3.6 \, \mu {\rm m}])>1.0$ and all of the 17 objects satisfying $z \lsim 0.5$ and $(K_s-[4.5 \, \mu {\rm m}])>1.5$ are very low mass systems, with assembled stellar masses $M<2.5 \times 10^9 \, M_\odot$. The majority of them are best-fit by young-galaxy templates.  The excess of near-IR colour could be explained by the presence of an active nucleus, whose IRAC magnitudes could not be properly  predicted using the GISSEL98 stellar templates.  On the other hand, an alternative and perhaps more likely explanation is that these very red $K_s$/IRAC colours are produced by the presence of prominent emission lines in the IRAC bands. The presence of an aromatic feature at $3.3 \, \mu {\rm m}$ is characteristic of star-forming galaxies. These aromatic features are not incorporated in the GISSEL98 templates, explaining why such very red colours could not have been predicted from the SED models. If this is the case, the colour excess would be the signature of intense star-formation in these young objects. Only near-IR spectrocopy would be able to confirm the nature of these sources.

\begin{figure*}
\begin{center}
\includegraphics[width=1.0\hsize,angle=0] {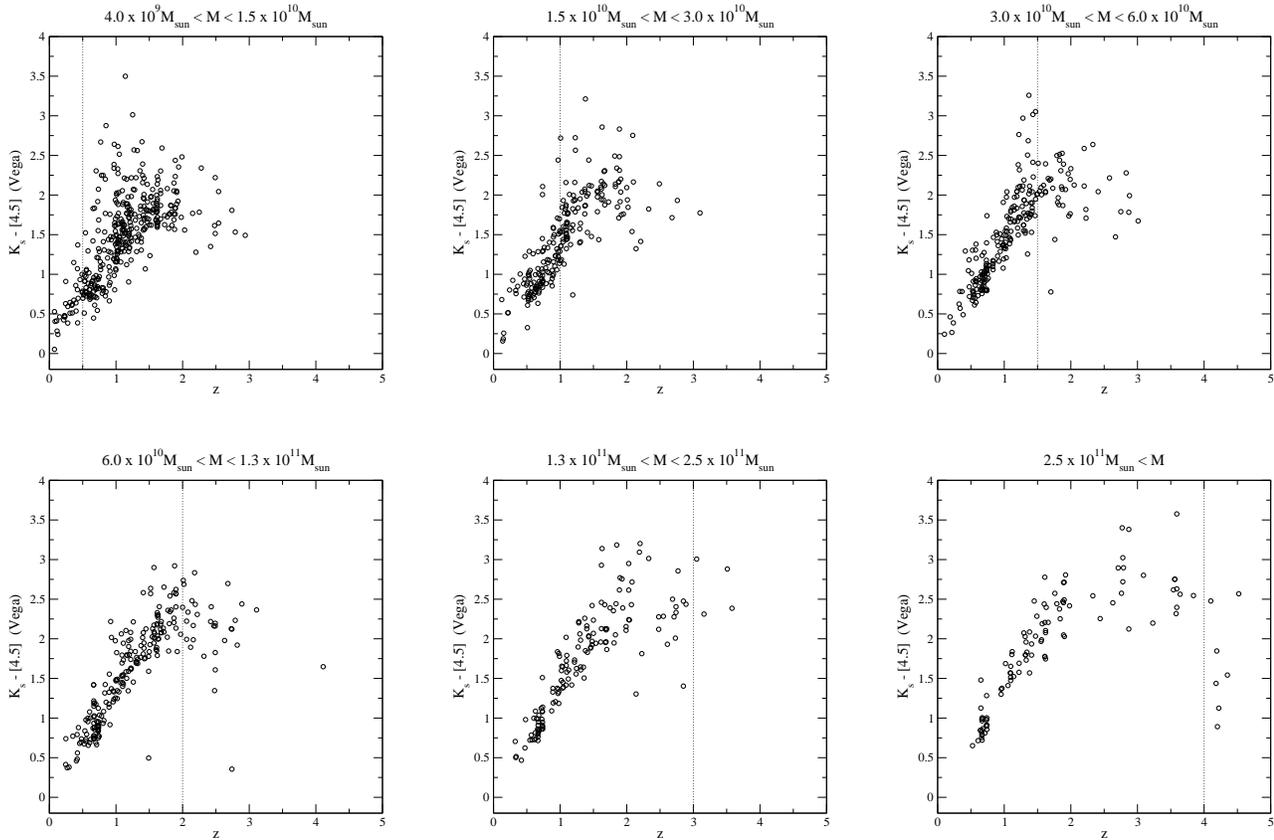}
\caption[]{\label{Ksm45comp} The observed $(K_s-[4.5 \, \mu \rm m])$ colours versus redshift for galaxies within different ranges of stellar mass. In each panel, the vertical dotted line delimits the region of completeness for each stellar mass bin.}
\end{center}  
\end{figure*}

    Figure~\ref{Ksm45comp} shows  the observed $(K_s-[4.5 \, \mu {\rm m}])$ colours versus redshift in different stellar mass bins, analogously to figure~\ref{imKmass}. We observe that the  $(K_s-[4.5 \, \mu {\rm m}])$ colours increase up to redshift $z \sim 2$, and the colour-redshift relation becomes tighter with increasing stellar mass. In contrast, the colour dispersion becomes quite large above that redshift, where the observed near-IR bands start to map the rest-frame optical bands. This colour dispersion is in part due to larger photometry uncertainties, but it is probably also an indication of a wide variety of star-formation histories and dust extinction in these galaxies.

\begin{figure}
\begin{center}
\includegraphics[width=1.0\hsize,angle=0] {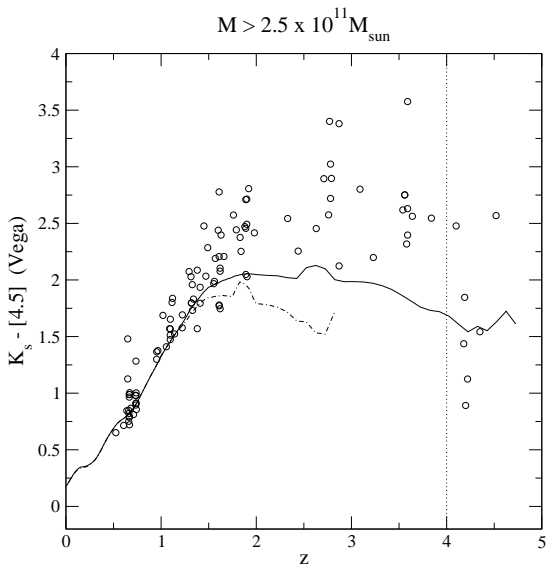}
\caption[]{\label{Ksm45model} The evolution of the observed $(K_s-[4.5 \, \mu \rm m])$ colours with redshift for galaxies with estimated stellar mass $M>2.5 \times 10^{11} \, M_\odot$. The solid and dotted-dashed lines show the modelled evolution of the $(K_s-[4.5 \, \mu \rm m])$ colours for galaxies formed in an instantaneous burst at redshift $z_f=5$ and $z_f=3$, respectively, both with no dust and  passive evolution thereafter.}
\end{center}  
\end{figure}

   Figure~\ref{Ksm45model} shows an amplified version of the high-mass panel in figure~\ref{Ksm45comp}. We added for comparison the modelled evolution of the observed $(K_s-[4.5 \, \mu {\rm m}])$ colours for galaxies formed in a single burst at redshifts $z_f=3$ and $z_f=5$, both with no dust and passive evolution thereafter. Some interesting conclusions can be extracted from the analysis of this figure. Firstly, it is quite evident that the observed near-IR colours of many massive galaxies at high redshifts ($z \gsim 1.5$) cannot be explained without considerable amounts of dust (cf. also Caputi et al. 2004). Independently of the formation redshift, the colours of a passively-evolving single burst are not sufficient to reproduce the red observer-frame near-IR colours of  massive galaxies. We conclude, then, that purely passive galaxies should be a minority at high redshifts.  Secondly, the colours observed at lower redshifts ($z \lsim 1.5$) indicate that the near-IR light of $M>2.5 \times 10^{11} \, M_\odot$ galaxies must be dominated by evolved stellar populations. A similar conclusion  has been obtained from the  analysis  of the  $(I_{775}-K_s)$ colours of these systems. 
   
   We note that the tight colour-redshift relation observed for high-mass galaxies at  $z \lsim 1.5-2.0$ does not imply a common  formation redshift or star-formation history. For instance, as we mentioned in Section~\ref{sec_masscol}, some of these galaxies are bright mid-IR sources, while others are not, suggesting a wide variety of on-going star-formation rates (Caputi et al.~2005b,c). However, and in contrast to what is progressively observed in less massive systems, all the plausible new star-formation in very massive galaxies at low redshifts could have a minor impact on their overall properties and, in particular, on their SED colours.

\section{Summary and concluding remarks}
\label{sec_conc}

\parskip=0pt

  With the GOODS project close to completion,  in this work we have been able to set tighter  constraints on  the evolution of near-IR-selected galaxies in the GOODS/CDFS field. We took advantage of the latest available very deep and homogeneous multiwavelength photometry, in conjunction with the information provided by complementary spectroscopic programs and the COMBO17 campaign, to construct an optimised redshift catalogue for the  $K_s<21.5$ galaxies in the GOODS-South field.    Based on this catalogue, we performed a detailed analysis of galaxy luminosity and mass evolution up to very high redshifts $z \sim 3-4$.

  The incorporation of the Spitzer/IRAC data at $3.6$ and $4.5\, \mu {\rm m}$  allowed us to compute rest-frame $K_s$-band luminosities and stellar masses with considerably smaller extrapolations than those applied in Caputi et al. (2005a). In spite of this important difference, we note that most of the main conclusions obtained by Caputi et al. (2005a) have been confirmed by this work. The addition of IRAC data did not produce any overall change in the derived galaxy luminosities and stellar masses. Other authors also have concluded that  the use of IRAC data produced a minor impact on the derived galaxy stellar masses, when other shorter optical/near-IR multiwavelength data are also available (e.g. Shapley et al. 2005).

  From the modelling of the evolution of the $K_s$-band luminosities, we determined that the  characteristic magnitude $M^\ast$ of the rest-frame $K_s$-band LF significantly brightens from redshifts $z=0$ to $\langle z \rangle=2.5$. This is simultaneously produced  with a decrease in the total density of objects, so we conclude  that the $K_s$-band LF has both a substantial luminosity and density evolution with redshift.  The characteristics of this evolution reveal some important aspects of the sequence of galaxy formation. For instance, that most of the brightest galaxies are assembled at high redshifts ($z \gsim 1$), while the creation of a substantial number of intermediate-to-low  luminosity objects might continue down to low redshifts (although the latter will need to be confirmed by future deeper surveys, which have direct access to low-luminosity objects at high redshifts).

  We have also studied the very high stellar-mass end ($\rm M>2.5 \times 10^{11} M_\odot$) of galaxy evolution. We found that most of these very massive systems in our sample have been formed in the $\rm \sim 3.5 \, Gyr$-period of time  between redshifts $z \approx 3$ and  $z \approx 1$, although a significant fraction (up to $\sim 20$\%) appear to have been in place before redshift $z=4$. Very recently, from the cross-correlation of our $K_s<21.5$ galaxy sample with the Spitzer/$\rm 24 \, \mu m$ galaxies  in the same field,  Caputi et al.~(2005b) found that the most-intense star-forming galaxies at redshifts $z \approx 2.0-3.0$ are mainly massive ($\rm M>1.5 \times 10^{11} M_\odot$) galaxies. This is also the main epoch of flourish of submillimetre galaxies (Chapman et al. 2005), which presumably are progenitors of the most massive local galaxies (e.g. Stevens et al. 2003). Thus, this period appears to be critical to understand the formation and evolution of the bulk of the most massive systems present in the local Universe.

   Theoretical models of galaxy formation in their  current status are  not completely able to account for all of these results. Several different  clues have been proposed in  recent works to try to improve the agreement of the models with the data at high redshifts, including e.g. the effects of QSO and supernova feedback (Granato et al. 2004; Cirasuolo et al. 2005), the adoption of an alternative IMF (Baugh et al. 2005) or the use of hydrodynamical models (Nagamine et al. 2005).    However, the discrepancies are not produced by the specific recipes used by theoretical models to account for star formation, but seem to lie at the more fundamental level of hierarchical model predictions. Hierarchical models do not seem to be able to reproduce the whole of observational properties from the local Universe to high redshifts.
  A likely alternative to this problem is the invalidity of a universal IMF when deriving stellar masses from the observed luminosities.   A definitive comprehension of the main epoch of very massive galaxy formation, as well as the determination of the initial distribution of stars in these systems, still  constitutes a major challenge from both the observational and theoretical points of view.


\section*{Acknowledgements}
This paper is based on observations made with the Advanced Camera for Surveys on board the Hubble Space Telescope operated by NASA/ESA; with the Infrared Spectrometer and Array Camera on the `Antu' Very Large Telescope operated by the European Southern Observatory in Cerro Paranal, Chile; and the Infrared Array Camera on board the Spitzer Space Telescope, operated by the Jet Propulsion Laboratory, California Institute of Technology, under a contract with NASA;  and  form part of the publicly available GOODS datasets.  We thank the GOODS teams for providing
reduced data products.

We are very grateful to Kentaro Nagamine for kindly providing us with his SPH simulation predictions, and to Will Percival for providing us the code to compute the number densities of dark matter haloes. We thank the anonymous referee for his/her useful comments and suggestions, which improved the discussion of results in this work. KIC acknowledges CNES and CNRS funding. RJM acknowledges the support of the Royal Society.


\bibliographystyle{mn2e}

\section*{References}

\bib Alonso-Herrero A. et al., 2005, ApJ, submitted

\bib Arnouts S. et al., 2001, A\&A, 379, 740

\bib Baugh C.M.,  Lacey C.G., Frenk C.S., Granato G.L., Silva L., Bressan A., Benson A.J., Cole S., 2005, MNRAS, 356, 1191

\bib Bell E.F., McIntosh D.H., Katz N., Weinberg M.D., 2003, ApJS, 149, 289

\bib Ben\'{\i}tez N., 2000, ApJ, 536, 571

\bib Bertin E., Arnouts S., 1996, A\&A, 117, 393

\bib Blakeslee J.P. et al., 2003, ApJ, 596, L143

\bib Bolzonella M., Miralles J.-M., Pell\'o R., 2000, A\&A, 363, 476

\bib Bond J. R., Cole S., Efstathiou G., Kaiser, N., 1991, ApJ, 379, 440 

\bib Bower  R. G., Lucey J.R., Ellis R.S., 1992, MNRAS, 254, 601

\bib Bruzual A. G., Charlot S., 1993, ApJ, 405, 538

\bib Bruzual A. G., Charlot S., 2003, MNRAS, 344, 1000

\bib Calzetti D., Armus L., Bohlin R. C., Kinney A. L., Koornneef
J., Storchi-Bergmann T., 2000, ApJ, 533, 682

\bib Caputi K. I., Dunlop J.S., McLure R.J., Roche N.D., 2004, MNRAS, 353, 30 

\bib Caputi K. I., Dunlop J.S., McLure R.J., Roche N.D., 2005a, MNRAS, 361, 607

\bib Caputi K. I. et al., 2005b, ApJ, in press (astro-ph/0510070)

\bib Caputi K. I. et al., 2005c, A\&A, submitted

\bib Chabrier G., 2003, PASP, 115, 763

\bib Chapman S.C, Blain A.W., Smail I.R., Ivison R.J., 2005, ApJ, 622, 772

\bib Cimatti A. et al., 2002, A\&A, 391, L1

\bib Cimatti A. et al., 2002b, A\&A, 392, 395

\bib Cimatti A. et al., 2004, Nature, 430, 184

\bib Cirasuolo M., Shankar F., Granato G.L., De Zotti G., Danese L., 2005, ApJ, submitted

\bib Cole S. et al., 2001, MNRAS, 326, 255

\bib Condon J.J. et al., 2003, AJ, 125, 2411 

\bib Conselice C.J., 2005, ApJ, submitted

\bib Daddi E. et al., 2003, ApJ, 588, 50

\bib Daddi E. et al., 2005, ApJ, 626, 680

\bib Dahlen T., Mobasher B., Somerville R.S., Moustakas L.A., Dickinson M., Ferguson H.C., Giavalisco M., 2005, ApJ, in press

\bib Dickinson M., Papovich C., Ferguson H., Budav\'ari T., 2003, ApJ, 587, 25

\bib Drory N., Bender R., Feulner G., Hopp U., Maraston C., Snigula J., Hill G.J., 2003, ApJ, 595, 698

\bib Drory N., Salvato M., Gabasch A., Bender R., Hopp U., Feulner G., Pannella M., 2005, ApJ, 619, L131

\bib Dunlop, J.S., 2005, in {\em Starbursts: From 30 Doradus to Lyman Break Galaxies},
 de Grijs R. and Gonz\'alez-Delgado R.M. eds., Astrophysics and Space Science Library, 329, 
 Dodrecht: Springer, 121

\bib  Ellis R.S., Smail I., Dressler A., Couch W.J., Oemler A. Jr., Butcher H., Sharples R.M., 1997, ApJ, 483, 582

\bib Fazio G. G. et al., 2004, ApJS, 154, 10

\bib Feulner G., Bender R., Drory N., Hopp U., Snigula J., Hill G. J., 2003, MNRAS, 342, 605

\bib Fontana A. et al., 2003, ApJ, 594, L9

\bib Fontana A. et al., 2004, A\&A, 424, 23

\bib Genzel R., Baker A.J., Tacconi L.J., Lutz D., Cox P., Guilloteau S., Omont A., 2003, ApJ, 584, 633

\bib Giavalisco M., the GOODS team, 2004, ApJ, 600, L93

\bib Glazebrook K. et al., 2004,  Nature, 430, 181

\bib Granato G.L., De Zotti G., Silva L., Bressan A., Danese L., 2004, ApJ, 600, 580

\bib Heavens A., Panter B., Jim\'{e}nez R., Dunlop J., 2004, Nature, 428, 625

\bib Kochanek C.S. et al., 2001, ApJ, 560, 566

\bib Kroupa P. \& Weidner C., 2005, to appear in {\em Massive Star Birth: A Crossroads of Astrophysics}, Cesaroni R., Churchwell E., Felli M. \& Walmsley C.M., eds., Proceedings IAU Symposium No.227 

\bib Le F\`evre O. et al., 2004, A\&A, 428, 1043

\bib Marshall H.L., Tananbaum H., Avni Y., Zamorani G., 1983, ApJ, 269, 35

\bib Martin D.C. et al., 2005, ApJ, 619, L1

\bib McCarthy P.J. et al., 2004, ApJ, 614, L9

\bib Nagamine K., Cen R., Hernquist L., Ostriker J.P., Springel V., 2005, ApJ, 627, 608

\bib Papovich C. et al., 2005, ApJ, submitted

\bib Patton D.R. et al., 2002, ApJ, 565, 208

\bib Poggianti B.M., 1997, A\&AS, 122, 399

\bib Pozzetti L. et al., 2003, A\&A, 402, 837

\bib Roche N.D., Dunlop J.S., Almaini O., 2003, MNRAS, 346, 803 

\bib Rudnick G. et al., 2003, ApJ, 599, 847

\bib Salpeter E.E., 1955, ApJ, 121, 161

\bib Saracco P. et al., 2005, MNRAS, 357, L40

\bib Shapley A.E., Steidel C.C., Erb D.K., Reddy N.A., Adelberger K.L., Pettini M., Barmby P., Huang J., 2005, ApJ, 626, 698

\bib Shechter P., 1976, ApJ, 203, 297

\bib Somerville R.S., Lee K., Ferguson H.C., Gardner J.P., Moustakas L.A., Giavalisco M., 2004, ApJ, 600, L171 

\bib Stanford S.A., Eisenhardt P.R., Dickinson M., 1998, ApJ, 492, 461

\bib Stevens J.A. et al., 2003, Nature, 425, 264

\bib Szokoly G.P. et al., 2004, ApJS, 155, 271 

\bib Vanzella E. et al., 2005, A\&A, 434, 53

\bib Werner M.W. et al., 2004,  ApJS, 154, 1

\bib Willott C.J., Rawlings S., Jarvis M.J., Blundell K.M.,
2003, MNRAS, 339, 173

\bib Wolf C. et al., 2004, A\&A, 421, 913

\end{document}